\documentclass[longauth]{aa/aa}

\usepackage{graphicx}
\usepackage{txfonts}
\usepackage{lipsum}
\usepackage{subcaption}         
\usepackage{lscape}             
\usepackage{longtable}
\usepackage{placeins}           

\usepackage{CJKutf8}     
\usepackage{tabularx}
\usepackage{threeparttable}
\usepackage{siunitx}

\def\gridline#1{\vskip6pt\hbox to\hsize{#1}\vskip6pt}

\def\fig#1#2#3{\hfill\vbox{\parskip=0pt\hsize=#2
\includegraphics[width=#2]{#1}\vskip2pt\vtop{\centering
\footnotesize
\hsize=#2
#3\vskip1pt
}}\hfill}

\newcommand*{\orcid}[1]{
    \href{https://orcid.org/#1}{\,\raisebox{0.2em}{
    \hspace{-1em}
    \includegraphics[height=0.6em,width=0.6em]{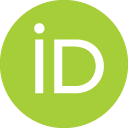}
}}}

\newcommand{\micron}{$\mu$m\xspace}
\newcommand{\solarmass}{M$_{\odot}$\xspace}
\newcommand{\velu}{km s$^{-1}$\xspace}

\usepackage[pdfpagelabels=false]{hyperref}      
\hypersetup{colorlinks=true,linkcolor=blue,citecolor=blue,filecolor=blue,urlcolor=blue,}

\begin{document} 

   \title{Quantifying the decline in CO(2–1) intensity around PHANGS-HST star clusters: Constraints on molecular gas dispersal, stellar feedback, and cluster formation efficiency}
    \titlerunning{Molecular gas around star clusters}


\newcommand{\OSU}{\label{OSU} Department of Astronomy, The Ohio State University, 140 West 18th Avenue, Columbus, Ohio 43210, USA}

\newcommand{\Alberta}{\label{Alberta} Department of Physics, University of Alberta, Edmonton, AB T6G 2E1, Canada}

\newcommand{\ANU}{\label{ANU} Research School of Astronomy and Astrophysics, Australian National University, Canberra, ACT 2611, Australia}

\newcommand{\IPAC}{\label{IPAC} Caltech-IPAC, 1200 E. California Blvd. Pasadena, CA 91125, USA}

\newcommand{\Carnegie}{\label{Carnegie} Observatories of the Carnegie Institution for Science, 813 Santa Barbara Street, Pasadena, CA 91101, USA}

\newcommand{\CCAPP}{\label{CCAPP} Center for Cosmology and Astroparticle Physics, 191 West Woodruff Avenue, Columbus, OH 43210, USA}

\newcommand{\CfA}{\label{CfA}Harvard-Smithsonian Center for Astrophysics, 60 Garden Street, Cambridge, MA 02138, USA}

\newcommand{\UCT}{\label{UCT}Department of Astronomy, University of Cape Town, Rondebosch 7701, Cape Town, South Africa}

\newcommand{\UCASS}{\label{UCASS}Department of Mathematical Sciences, Centre for Astrophysics and Space Science, University of South Africa, Florida, 1709, South Africa}

\newcommand{\CITEVA}{\label{CITEVA} Centro de Astronomía (CITEVA), Universidad de Antofagasta, Avenida Angamos 601, Antofagasta, Chile}

\newcommand{\CNRS}{\label{CNRS} CNRS, IRAP, 9 Av. du Colonel Roche, BP 44346, F-31028 Toulouse cedex 4, France}

\newcommand{\ESO}{\label{ESO} European Southern Observatory, Karl-Schwarzschild Stra{\ss}e 2, D-85748 Garching bei M\"{u}nchen, Germany}

\newcommand{\ESOChile}{\label{ESOChile} European Southern Observatory, Avenida Alonso de Cordoba 3107, Casilla 19, Santiago 19001, Chile}

\newcommand{\HD}{\label{HD} Astronomisches Rechen-Institut, Zentrum f\"{u}r Astronomie der Universit\"{a}t Heidelberg, M\"{o}nchhofstra\ss e 12-14, D-69120 Heidelberg, Germany}

\newcommand{\ICRAR}{\label{ICRAR} International Centre for Radio Astronomy Research, University of Western Australia, 35 Stirling Highway, Crawley, WA 6009, Australia}

\newcommand{\IRAM}{\label{IRAM} Institut de Radioastronomie Millim\'{e}trique (IRAM), 300 Rue de la Piscine, F-38406 Saint Martin d'H\`{e}res, France}

\newcommand{\ITA}{\label{ITA} Universit\"{a}t Heidelberg, Zentrum f\"{u}r Astronomie, Institut f\"{u}r Theoretische Astrophysik, Albert-Ueberle-Str 2, D-69120 Heidelberg, Germany}

\newcommand{\IWR}{\label{IWR} Universit\"{a}t Heidelberg, Interdisziplin\"{a}res Zentrum f\"{u}r Wissenschaftliches Rechnen, Im Neuenheimer Feld 205, D-69120 Heidelberg, Germany}

\newcommand{\JHU}{\label{JHU} Department of Physics and Astronomy, The Johns Hopkins University, Baltimore, MD 21218, USA}

\newcommand{\Leiden}{\label{Leiden} Leiden Observatory, Leiden University, P.O. Box 9513, 2300 RA Leiden, The Netherlands}

\newcommand{\Maryland}{\label{Maryland} Department of Astronomy, University of Maryland, College Park, MD 20742, USA}

\newcommand{\MPE}{\label{MPE} Max-Planck-Institut f\"{u}r extraterrestrische Physik, Giessenbachstra{\ss}e 1, D-85748 Garching, Germany}

\newcommand{\MPIA}{\label{MPIA} Max-Planck-Institut f\"{u}r Astronomie, K\"{o}nigstuhl 17, D-69117, Heidelberg, Germany}

\newcommand{\Nagoya}{\label{Nagoya} Department of Physics, Nagoya University, Furo-cho, Chikusa-ku, Nagoya, Aichi 464-8602, Japan}

\newcommand{\NRAO}{\label{NRAO} National Radio Astronomy Observatory, 520 Edgemont Road, Charlottesville, VA 22903-2475, USA}

\newcommand{\OAN}{\label{OAN} Observatorio Astron\'{o}mico Nacional (IGN), C/Alfonso XII, 3, E-28014 Madrid, Spain}

\newcommand{\ObsParis}{\label{ObsParis} Sorbonne Universit\'{e}, Observatoire de Paris, Universit\'{e} PSL, CNRS, LERMA, F-75014, Paris, France}

\newcommand{\Princeton}{\label{Princeton} Department of Astrophysical Sciences, Princeton University, Princeton, NJ 08544 USA}

\newcommand{\UToledo}{\label{UToledo} University of Toledo, 2801 W. Bancroft St., Mail Stop 111, Toledo, OH, 43606}

\newcommand{\Toulouse}{\label{Toulouse} Universit\'{e} de Toulouse, UPS-OMP, IRAP, F-31028 Toulouse cedex 4, France}

\newcommand{\UBonn}{\label{UBonn} Argelander-Institut f\"ur Astronomie, Universit\"at Bonn, Auf dem H\"ugel 71, 53121 Bonn, Germany}

\newcommand{\UChile}{\label{UChile} Departamento de Astronom\'{i}a, Universidad de Chile, Camino del Observatorio 1515, Las Condes, Santiago, Chile}

\newcommand{\UConn}{\label{UConn} Department of Physics, University of Connecticut, Storrs, CT, 06269, USA}

\newcommand{\UCSD}{\label{UCSD}Center for Astrophysics and Space Sciences, Department of Physics,  University of California, San Diego, 9500 Gilman Drive, La Jolla, CA 92093, USA}

\newcommand{\UGent}{\label{UGent} Sterrenkundig Observatorium, Universiteit Gent, Krijgslaan 281 S9, B-9000 Gent, Belgium}

\newcommand{\ULyon}{\label{ULyon} Univ Lyon, Univ Lyon 1, ENS de Lyon, CNRS, Centre de Recherche Astrophysique de Lyon UMR5574,\\ F-69230 Saint-Genis-Laval, France}

\newcommand{\UMass}{\label{UMass} University of Massachusetts—Amherst, 710 N. Pleasant Street, Amherst, MA 01003, USA}

\newcommand{\UWyoming}{\label{UWyoming} Department of Physics and Astronomy, University of Wyoming, Laramie, WY 82071, USA}

\newcommand{\LAM}{\label{LAM} Aix Marseille Univ, CNRS, CNES, LAM (Laboratoire d’Astrophysique de Marseille), Marseille, France}

\newcommand{\UHawaii}{\label{UHawaii} Institute for Astronomy, University of Hawaii, 2680 Woodlawn Drive, Honolulu, HI 96822, USA}

\newcommand{\UCM}{\label{UCM} Departamento de F\'{\i}sica de la Tierra y Astrof\'{\i}sica, Universidad Complutense de Madrid, E-28040, Spain}

\newcommand{\IPARC}{\label{IPARC} Instituto de F\'{\i}sica de Part\'{\i}culas y del Cosmos IPARCOS, Facultad de Ciencias F\'{\i}sicas, Universidad Complutense de Madrid, E-28040, Spain}

\newcommand{\STScI}{\label{STScI} Space Telescope Science Institute, 3700 San Martin Drive, Baltimore, MD 21218, USA}

\newcommand{\McMaster}{\label{McMaster} Department of Physics and Astronomy, McMaster University, 1280 Main Street West, Hamilton, ON L8S 4M1, Canada}

\newcommand{\INAF}{\label{INAF} INAF -- Osservatorio Astrofisico di Arcetri, Largo E. Fermi 5, I-50157, Firenze, Italy}

\newcommand{\Sydney}{\label{Sydney} Sydney Institute for Astronomy, School of Physics A28, The University of Sydney, NSW 2006, Australia}

\newcommand{\CITA}{\label{CITA} Canadian Institute for Theoretical Astrophysics (CITA), University of Toronto, 60 St George St, Toronto, ON M5S 3H8, Canada}

\newcommand{\ASIAA}{\label{ASIAA} Institute of Astronomy and Astrophysics, Academia Sinica, No. 1, Sec. 4, Roosevelt Road, Taipei 10617, Taiwan}

\newcommand{\TKU}{\label{TKU} Department of Physics, Tamkang University, No.151, Yingzhuan Rd., Tamsui Dist., New Taipei City 251301, Taiwan}

\newcommand{\PSMA}{\label{PSMA} Penn State Mont Alto, 1 Campus Drive, Mont Alto, PA  17237, USA}

\newcommand{\ILL}{\label{ILL} ILL}

\newcommand{\stromlo}{\label{stromlo} Research School of Astronomy and Astrophysics, Australian National University, Mt Stromlo Observatory, Weston Creek, ACT 2611, Australia}

\newcommand{\UCatolica}{\label{UCatolica} Instituto de Astronom\'ia, Universidad Cat\'olica del Norte, Av. Angamos 0610, Antofagasta, Chile}

\newcommand{\UT}{\label{UT} McDonald Observatory, The University of Texas at Austin, 1 University Station, Austin, TX 78712-0259, USA}

\newcommand{\Vanderbilt}{\label{Vanderbilt} Department of Physics and Astronomy, Vanderbilt University, VU Station 1807, Nashville, TN 37235, USA}

\newcommand{\UNF}{\label{UNF} Department of Physics, University of North Florida, 1 UNF Dr. Jacksonville FL 32224}

\newcommand{\NAOC}{\label{NAOC} Chinese Academy of Sciences South America Center for Astronomy, National Astronomical Observatories, CAS, Beijing 100101, China}

\newcommand{\CASA}{\label{CASA} Center for Astrophysics and Space Astronomy, University of Colorado, 389 UCB, Boulder, CO 80309-0389, USA}

\newcommand{\UNAM}{\label{UNAM} Universidad Nacional Aut\'onoma de M\'exico, Instituto de Astronom\'ia, AP 106, Ensenada 22800, BC, M\'exico}

\newcommand{\UDP}{\label{UDP} Instituto de Estudios Astrof\'isicos, Facultad de Ingenier\'ia y Ciencias, Universidad Diego Portales, Av. Ej\'ercito Libertador 441, Santiago, Chile}

\newcommand{\Steward}{\label{Steward} Steward Observatory, University of Arizona, 933 N. Cherry Ave., Tucson, AZ 85721-0065, USA}  

\newcommand{\APO}{\label{APO} Apache Point Observatory and New Mexico State University, P.O.\ Box 59,
Sunspot, NM 88349-0059, USA}

\newcommand{\UNAMCU}{\label{UNAMCU} Universidad Nacional Aut\'onoma de M\'exico, Instituto de Astronom\'ia, AP 70-264, CDMX 04510, M\'exico}

\newcommand{\UWash}{\label{UWash}Department of Astronomy, University of Washington, Seattle, WA, 98195}

\newcommand{\CC}{\label{CC}Department of Physics, Colorado College, Colorado Springs, CO 80903}

\newcommand{\Utah}{\label{Utah}Department of Physics and Astronomy, University of Utah, 115 S. 1400 E., Salt Lake City, UT 84112, USA}

\newcommand{\UConcepcion}{\label{UConcepcion}Departamento de Astronom\'ia, Universidad de Concepci\'on, Casilla 160-C, Concepci\'on, Chile}

\newcommand{\FCLA}{\label{FCLA}Franco-Chilean Laboratory for Astronomy, IRL 3386, CNRS and Universidad de Chile, Santiago, Chile}

\newcommand{\Oklahoma}{\label{Oklahoma}Homer L. Dodge Department of Physics and Astronomy, University of Oklahoma, Norman, OK 73019, USA}

\newcommand{\UIUC}{\label{UIUC}Department of Astronomy, University of Illinois, Urbana, IL 61801, USA}

\newcommand{\Harvard}{\label{Harvard}Harvard-Smithsonian Center for Astrophysics, Cambridge, MA 02138, USA}

\newcommand{\caltech}{\label{caltech}Department of Astronomy, California Institute of Technology, Pasadena, CA 91125, USA}

\newcommand{\UOA}{\label{UOA}Department of Physics, University of Arkansas, 226 Physics Building, 825 West Dickson Street, Fayetteville, AR 72701, USA}

\newcommand{\units}{\label{units}Department of Physics, Astronomy Section, University of Trieste, Via G.B. Tiepolo, 11, I-34143 Trieste, Italy}

\newcommand{\Rad}{\label{Rad}{Elizabeth S. and Richard M. Cashin Fellow at the Radcliffe Institute for Advanced Studies at Harvard University, 10 Garden Street, Cambridge, MA 02138, USA}}

\newcommand{\LJMU}{\label{LJMU}{Astrophysics Research Institute, Liverpool John Moores University, IC2, Liverpool Science Park, 146 Brownlow Hill, Liverpool L3 5RF, UK}}

\newcommand{\COOL}{\label{COOL}{Cosmic Origins Of Life (COOL) Research DAO, \href{https://coolresearch.io}{https://coolresearch.io}}}

\newcommand{\Ox}{\label{Ox}{Sub-department of Astrophysics, Department of Physics, University of Oxford, Keble Road, Oxford OX1 3RH, UK}}

\newcommand{\UShizuokaGlobal}{\label{UShizuokaGlobal}{Faculty of Global Interdisciplinary Science and Innovation, Shizuoka University, 836 Ohya, Suruga-ku, Shizuoka 422-8529, Japan}}

\newcommand{\NAOJ}{\label{NAOJ}{National Astronomical Observatory of Japan, 2-21-1 Osawa, Mitaka, Tokyo, 181-8588, Japan}}

\newcommand{\StU}{\label{StU}{SUPA, School of Physics and Astronomy, University of St Andrews, North Haugh, St Andrews, KY16 9SS}}

\newcommand{\JBCA}{\label{JBCA}{UK ALMA Regional Centre Node, Jodrell Bank Centre for Astrophysics, Department of Physics and Astronomy, The University of Manchester, Oxford Road, Manchester M13 9PL, UK}}

\newcommand{\UniCA}{\label{UniCA}{Université Côte d'Azur, Observatoire de la Côte d'Azur, CNRS, Laboratoire Lagrange, 06000, Nice, France}}

\newcommand{\IALP}{\label{IALP} Instituto de Astrofísica de La Plata, CONICET-UNLP, Paseo del Bosque S/N, B1900FWA La Plata, Argentina }

\newcommand{\UVA}{\label{UVA} Department of Astronomy, University of Virginia, Charlottesville, VA 22904, USA}

\newcommand{\Cambridge}{\label{Cambridge} Institute of Astronomy and Kavli Institute for Cosmology, University of Cambridge, Madingley Road, Cambridge, CB3 0HA, UK}

\newcommand{\UKY}{\label{UKY} Department of Physics and Astronomy, University of Kentucky, 506 Library Drive, Lexington, KY 40506, USA}

\newcommand{\Dyn}{\label{Dyn} Cluster of Excellence DYNAVERSE, Institute for Astrophysics, University of Cologne, Zülpicher Str. 77, 50937 Cologne, Germany}


\author{
       Hao He \begin{CJK*}{UTF8}{gbsn}(何浩)\end{CJK*}\inst{\ref{UBonn}, \ref{Dyn}} \thanks{\email{hhe@astro.uni-bonn.de}} \orcid{0000-0001-9020-1858}  
       \and  Adam K. Leroy\inst{\ref{OSU}, \ref{CCAPP}} \orcid{0000-0002-2545-1700} 
       \and Erik  Rosolowsky \inst{\ref{Alberta}} \orcid{0000-0002-5204-2259} 
       \and Janice Lee \inst{\ref{STScI}} \orcid{0000-0002-2278-9407}
       \and Brad Whitmore \inst{\ref{STScI}} \orcid{0000-0002-3784-7032}
       \and David Thilker \inst{\ref{JHU}} \orcid{0000-0002-8528-7340}
       \and Daniel~A.~Dale \inst{\ref{UWyoming}} \orcid{0000-0002-5782-9093}
       \and Frank Bigiel \inst{\ref{UBonn}, \ref{Dyn}} \orcid{0000-0003-0166-9745}
       \and Annie Hughes \inst{\ref{Toulouse}} \orcid{0000-0002-9181-1161} 
       \and Jiayi~Sun \begin{CJK*}{UTF8}{gbsn}(孙嘉懿)\end{CJK*}\inst{\ref{UKY}}\orcid{0000-0003-0378-4667}
       \and Gagandeep Anand\inst{\ref{STScI}} \orcid{0000-0002-5259-2314}
       \and Ashley~T.~Barnes \inst{\ref{ESO}} \orcid{0000-0003-0410-4504}
        \and Zein Bazzi \inst{\ref{UBonn}}\orcid{0009-0001-1221-0975}
       \and Médéric Boquien\inst{\ref{UniCA}}\orcid{0000-0003-0946-6176}
       \and Rupali Chandar \inst{\ref{UToledo}}\orcid{0000-0003-0085-4623}
       \and Dario Colombo\inst{\ref{UBonn}} \orcid{0000-0001-6498-2945}
       \and Sinan Deger \inst{\ref{Cambridge}} \orcid{0000-0003-1943-723X}
      \and Simthembile Dlamini \inst{\ref{UCT}, \ref{UCASS}}\orcid{0000-0002-2885-6172}
        \and Hamid Hassani \inst{\ref{Alberta}}\orcid{0000-0002-8806-6308}
        \and Stephan Hannon\inst{\ref{MPIA}}\orcid{/0000-0001-9628-8958}
       \and Ralf S. Klessen \inst{\ref{ITA},\ref{IWR}} \orcid{0000-0002-0560-3172} 
       \and Simon C. O. Glover \inst{\ref{ITA}} \orcid{0000-0001-6708-1317}
       \and Kathryn Grasha \inst{\ref{ANU}} \orcid{0000-0002-3247-5321}
        \and Ivan Grasimov \inst{\ref{UniCA}} \orcid{0000-0001-7113-8152}
        \and Remy Indebetouw\inst{\ref{UVA}, \ref{NRAO}} \orcid{0000-0002-4663-6827}
       \and Kirsten Larson \inst{\ref{STScI}} \orcid{0000-0003-3917-6460}
        \and Lukas Neumann \inst{\ref{ESO}} \orcid{0000-0001-9793-6400}
        \and Elias K. Oakes \inst{\ref{UConn}}\orcid{0000-0002-0119-1115}
        \and Hsi-An Pan \inst{\ref{TKU}}\orcid{0000-0002-1370-6964}
        \and Miguel Querejeta \inst{\ref{OAN}} \orcid{0000-0002-0472-1011}
       \and Lise Ramambason\inst{\ref{ITA}}\orcid{0000-0002-9190-9986}
       \and M. Jimena Rodríguez\inst{\ref{STScI}, \ref{IALP}} 
       \orcid{0000-0002-0579-6613}
        \and Sumit K. Sarbadhicary \inst{\ref{JHU}}\orcid{0000-0002-4781-7291}
       \and Eva Schinnerer \inst{\ref{MPIA}}\orcid{0000-0002-3933-7677}
       \and Qiushi (Chris) Tian \inst{\ref{Leiden}} \orcid{/0009-0009-9148-2159}
       \and Leonardo \'Ubeda \inst{\ref{STScI}} \orcid{0000-0001-7130-2880}
       \and Antonio Usero \inst{\ref{OAN}} \orcid{0000-0003-1242-505X}
       \and Thomas G. Williams\inst{\ref{JBCA}}\orcid{0000-0002-0012-2142}
       \and Jianwen Zhou \inst{\ref{UniCA}}\orcid{0009-0004-2927-239X}
}

\institute{\tiny
\UBonn     \and 
\Dyn \and 
\OSU \and 
\CCAPP  \and
\Alberta \and
\STScI \and
\JHU \and
\UWyoming \and
\Toulouse \and
\UKY \and 
\ESO \and 
\UniCA \and
\UToledo \and
\Cambridge \and
\UCT \and
\UCASS \and
\MPIA \and
\ITA \and
\IWR \and
\ANU \and
\UVA \and
\NRAO \and
\UConn \and
\TKU \and 
 \OAN  \and
\IALP \and
\Leiden \and
\JBCA
}

   \date{Received September 30, 20XX}

 
  \abstract
   {Star clusters form from molecular clouds. The mass of the stars formed depends on the molecular gas mass and the cluster formation efficiency. Over time, stellar feedback and cluster drift lead star clusters to become dis-associated from their parent clouds. 
   Previous studies have used cross-correlation functions and nearest-neighbour distributions to characterize this relationship. While these spatial statistics provide important measures of the association between star clusters and molecular gas, they do not by themselves quantify the surrounding molecular gas mass or its relation to cluster mass.
   }
   {
   We combine the largest survey of star clusters in nearby galaxies to-date, PHANGS-HST, with maps of molecular gas maps from PHANGS--ALMA to assess how the molecular gas associated with clusters depends on their age and mass. We aim to use these measurements to constrain the efficiency of cluster formation and the typical timescales over which clusters become dis-associated from their natal molecular gas. 
   }
   {We analyse 29,233 star clusters in 36 galaxies catalogued from PHANGS-HST galaxies and compare them to CO(2-1) maps at 150~pc resolution from PHANGS--ALMA. For each star cluster with stellar mass $\geq 10^{3.5}$~M$_\odot$, we measure the profile of line-integrated CO (2-1) intensity. To do this, we subtract the large-scale background to obtain the local background-subtracted CO intensity, $I_{\rm CO}^{\rm bkgsub}$. We group the clusters by age and mass inferred from spectral energy distribution modelling to quantify the dependence of $I_{\rm CO}^{\rm bkgsub}$ and corresponding molecular gas mass, $M_{\rm mol}$, on these cluster properties. To account for the cluster mass ($M_{\rm SC}$) dependence, we also calculate the normalised mass ratio of $(M_{\rm mol}/M_{\rm SC})$ for clusters with different ages. 
   }
   {CO intensity varies systematically with cluster age and mass. Clusters that are spatially associated with H$\alpha$ emission 
   exhibit the highest $I_{\rm CO}^{\rm bkgsub}$ of 7.4 K~km~s$^{-1}$, corresponding to a $M_{\rm mol}/M_{\rm SC}$ ratio of 82 and implying a cluster formation efficiency ($\epsilon_{\rm obs}$) of $\sim 1.2\%$. 
   Young clusters without significant associated H$\alpha$ emission show lower and age-dependent $I_{\rm CO}^{\rm bkgsub}$ and $M_{\rm mol}/M_{\rm SC}$, decreasing from 1.3 K~km~s$^{-1}$ and 15 for clusters with age 1--3 Myr, to a level consistent with background for clusters older than 9--10 Myr. Clusters older than 10~Myr show no statistical excess of CO intensity. We use the clusters with H$\alpha$ association as initial conditions to model the impact of cluster drift and demonstrate that this is unlikely to drive the observed cluster-gas spatial disassociation on short timescales of 4--6 Myr. Instead, early stellar feedback seems likely to be the main driver.
   }
  {}

   \keywords{galaxies: star clusters: general - galaxies: star formation - galaxies: ISM - ISM: structure - galaxies: spiral}

   \maketitle

\section{Introduction}

Massive stars produce ionising radiation, stellar winds, and supernova (SN) explosions. This stellar feedback shapes the interstellar medium (ISM), regulates star formation, contributes to ISM heating, drives turbulent motions, and creates galactic outflows \citep[][]{leitherer95, agertz13, Krumholz2014,Klessen2016,Girichidis2020}. Stars form from cold molecular gas and then exert this feedback on their surrounding birth clouds. 

Numerous observational studies that leverage H$\alpha$ or mid-infrared (mid-IR) emission to trace recent star formation have suggested short feedback timescales of $1{-}5$~Myr \citep[e.g.][]{kawamura_second_2009, corbelli_molecules_2017, hannon_h_2019, schinnerer_gas_2019, hannon22, chevance_life_2022, kim_environmental_2022, rodriguez23, rodriguez_star_2025,knutas_feast_2025, ramambason_duration_2026,linden_feast_2026}. This suggests rapid clearing of gas and dust by feedback before any stars explode as supernovae, which seems to be in agreement with predictions from some simulations \citep[e.g.][]{wainer_2026}. Rapid gas clearing has important implications, including leaving supernovae to explode in lower-density gas pre-processed by earlier stellar feedback, which also seems to be supported by observations \citep[e.g.][]{maykerchen_2023,sarbadhicary_2026}.

An important complementary approach is to measure the amount of cold molecular gas at the locations of young star clusters \citep[e.g.][]{kawamura_second_2009, whitmore_alma_2014, sun_hidden_2024}. Star clusters can be used as ``clocks'' for timing various astrophysical processes because their ages can be estimated from modelling their spectral energy distribution (SED) as simple stellar populations (SSPs) \citep[e.g.][]{larsen99, bc03, fall_age_2005, adamo17}.
By measuring the molecular gas content for clusters spanning a range of ages, it is possible to follow statistically how the association between clusters and their natal molecular gas changes as clusters age and evolve.

The PHANGS-HST survey \citep{lee_phangs-hst_2022} has identified and characterized $\sim 100,000$ star clusters and stellar associations \citep{maschmann_phangs-hst_2024,thilker_phangs-hst_2025} in galaxies where CO (2-1), tracing the molecular gas, has been mapped by the PHANGS--ALMA survey \citep{leroy_phangs-alma_2021}. In this paper we take advantage of this large, homogeneous dataset to measure the amount of CO emission at the locations of individual star clusters. We build on the aforementioned studies of individual galaxies or smaller samples, 
especially \citet{turner_phangs_2022}, who analysed 11 of these targets using an early version of the PHANGS-HST cluster catalogue.

Our approach focuses on measuring the amount and distribution of excess CO emission associated with clusters in different age, mass, or location bins. Physically, we aim to use clusters to infer the timescale over which cold gas is no longer associated with a young star cluster.  
Previous studies have quantified this association by identifying molecular clouds and assessing their spatial correspondence with young star clusters. For example, \citet[][]{grasha_connecting_2018,grasha_spatial_2019,turner_phangs_2022} measured cluster--cloud separations as a function of cluster age, and \citet[][]{grasha_spatial_2019,turner_phangs_2022} used spatial correlation functions to characterize the relative distributions of clouds and clusters. However, methods based on molecular cloud catalogues require the CO emission to be segmented into discrete structures, which has well-known biases and loses information near the resolution of the data \citep[e.g.][]{rosolowsky_giant_2021,he_structure_2026}.
The CO intensity field itself offers the most direct information on how the amount and spatial distribution of molecular gas around clusters change with cluster age. Therefore, rather than segmenting the CO emission and working with cloud catalogues, we follow \citet{he_structure_2026} in working with the intensity maps to measure the CO intensity field directly at and around the cluster locations.

Specifically, we measure the excess CO~(2-1) emission associated with each individual cluster after subtracting a local background. We then examine how CO content depends on cluster age and mass. By stacking measurements for large populations of clusters, this approach measures both the decline in CO intensity with cluster age and changes in the spatial distribution of the associated molecular gas.
This leverages the large available sample size and helps control for the significant resolution mismatch between the HST data used to identify the clusters ($\lesssim 10$~pc) and the CO data (common resolution $150$~pc). 

The paper is organised as follows: In Section \ref{sec:data}, we describe the star cluster catalogues and CO intensity maps used in our analyses. In Section \ref{sec:profiles}, we report measurements of CO (2-1) emission at the locations of star clusters, including background subtracted intensity measurements, the correlation between cluster mass and CO emission, and the stacked radial profile of CO emission around cluster locations, all binning clusters by mass and age. In Section \ref{sec:mock_control}, we construct a mock star cluster catalogue to explore the cause of the scatter in our obtained relation and certain features in our  measured radial profiles.
In Section \ref{sec:discussion}, we discuss the physical implications of our measured cluster-CO correlations for stellar feedback. We summarize our findings and conclusions in Section \ref{sec:conclusion}. 

\section{Data\label{sec:data}}

We analyse star clusters identified from the PHANGS-HST (GO-15654) survey \citep{lee_phangs-hst_2022}. These PHANGS-HST targets are a subset of 38 galaxies from the PHANGS-ALMA survey \citep{leroy_phangs-alma_2021}.
These are relatively massive, relatively face-on star-forming disk galaxies and the HST and ALMA observations cover most of the star formation activity and molecular gas in each target.
We use galaxy centres taken from \citet{querejeta_stellar_2021}, distances from \citet{anand_distances_2021, lee_phangs-hst_2022}, and inclination and orientations from \citet{leroy_phangs-alma_2021}. A total of 36 galaxies have PHANGS-HST cluster catalogues and PHANGS-ALMA CO map coverage at a common physical resolution of 150~pc; we omit NGC 1672 and NGC 4654, whose CO(2-1) maps have coarser resolution. 

\subsection{HST star cluster catalogue}

\begin{figure*}
    \centering
    \includegraphics[width=0.9\textwidth]{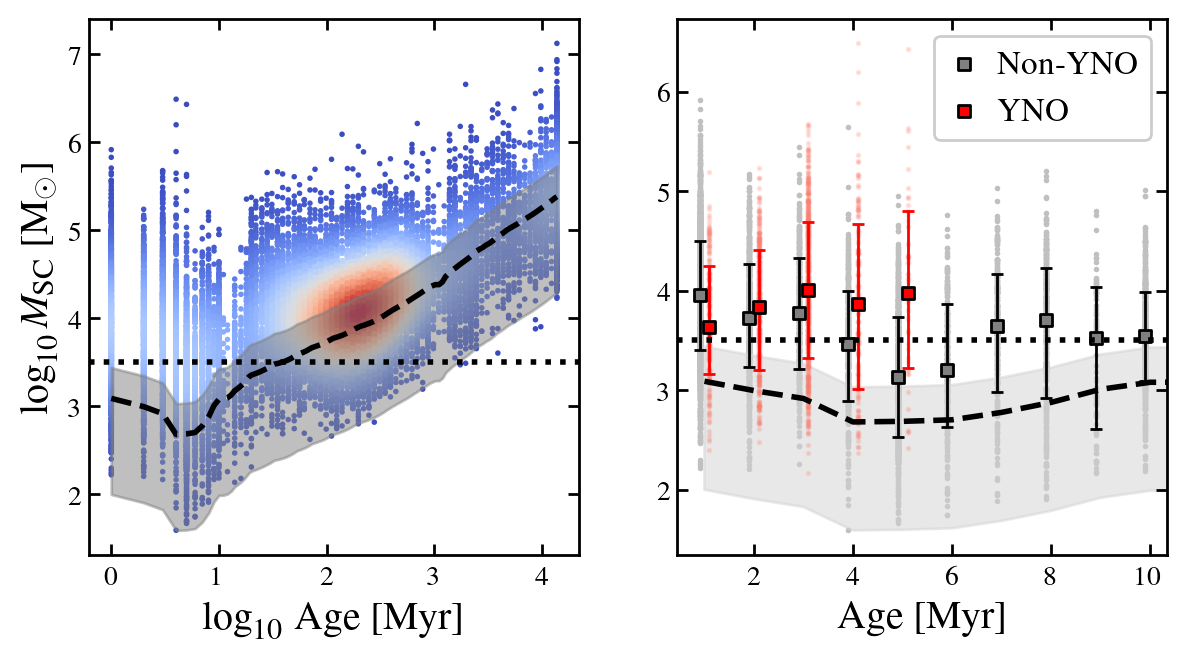}
    \caption{(\textit{Left}) Cluster mass versus age scatter plot for all clusters in our sample with points colour-coded by data density in this space. The dashed curve indicates the approximate cluster mass sensitivity as a function of age for a nominal detection limit of $V=24$ mag at 10 Mpc ($M_V=-6$ mag). The gray shaded region shows analogous curves for the range of distances in our galaxy sample. Although the true selection function is more complex, this curve has been used in the PHANGS-HST survey as a convenient reference \citep{lee_phangs-hst_2022}.    More sophisticated completeness modelling using artificial cluster injection-recovery \citep[e.g.][]{tang_cluster_2026} could refine this simple mass threshold in future work. The horizontal line indicates a constant mass of 10$^{3.5}$ \solarmass that we later adopt as a mass threshold. 
    (\textit{Right}) Zoomed-in version for clusters younger than 10 Myr. Gray dots indicate non-YNO clusters, and red dots indicate YNO clusters. Bold points and error bars shows the binned median and 16th-84th scatter.   }
    \label{fig:mass_vs_age}
\end{figure*}

\begin{table}[htb]
\caption{\label{tab:sample} HST cluster sample in 36 galaxies used in this work}
\centering
\renewcommand{\arraystretch}{1.4}
\begin{tabularx}{0.5\textwidth}{ccccc}
\hline\hline 
Sample & N$_{\mathrm{SC}}$ & $\log_{10}$ Age & $\log_{10}$ Mass & $E(B-V)$ \\
 & & [Myr] & [\solarmass] & \\
(1) & (2) & (3) & (4) & (5) \\
\hline
Total & 33483 & $2.0^{2.9}_{0.5}$ & $4.1^{4.8}_{3.5}$ & $0.1^{0.3}_{0.0}$ \\
Out of FOV$^{\rm a}$ & 3411 & $2.1^{2.9}_{0.5}$ & $3.7^{4.5}_{3.1}$ & $0.1^{0.3}_{0.0}$ \\
Center$^{\rm a}$ & 840 & $1.3^{3.5}_{0.5}$ & $5.0^{5.8}_{4.2}$ & $0.2^{0.6}_{0.1}$ \\
Selected & 29233 & $2.0^{2.8}_{0.5}$ & $4.1^{4.8}_{3.5}$ & $0.1^{0.3}_{0.0}$ \\
\hline
\end{tabularx}
\tablefoot{Columns: (1) Cluster sample.`Total': the full sample before selection; `Out of FOV': clusters outside the CO field of view (excluded); `Center': clusters in the central regions of galaxies (excluded); `Selected': sample that goes into our analyses. (2) The number of clusters in the corresponding sample. (3) The median age and the 16th--84th percentile range. (4) The median mass and the 16th-84th percentile range. (5) The median reddening and the 16th--84th percentile range. \\
Footnote: a. One cluster in NGC 6744 is both at the centre and out of FOV. }
\end{table}

We use the PHANGS-HST DR5 star cluster catalogues \citep{maschmann_phangs-hst_2024, thilker_phangs-hst_2025}. Clusters are identified and characterized using the images in five filters: F275W (\textit{NUV}), F336W (\textit{U}), F435W/F438W (\textit{B}), F555W (\textit{V}), and F814W (\textit{I}). Source detection is performed using DOLPHOT v2.0 \citep{dolphin_2016}  and DAOStarFinder on the \textit{V}-band images. Cluster candidates are selected from the detected sources based on multiple concentration indices \citep{Thilker2022}. These candidates are then classified into four categories based on either human classification \citep{whitmore_star_2021} or convolutional neural network models \citep{hannon_star_2023}: single peak sources with symmetric outer isophotes (C1), single peak sources with asymmetric outer isophotes (C2), sources with multiple peaks (C3), and other types of sources (e.g.\ artifacts, background galaxies and individual stars, C4). In this paper, we use candidate star clusters with classification C1 and C2 from the machine learning catalogue. The machine learning catalogue has a larger sample size and a higher completeness level than the human-classified catalogue \citep[][Table 3]{maschmann_phangs-hst_2024}.

The age, mass and reddening of clusters are derived using the spectral energy distribution (SED) fitting code \texttt{CIGALE} \citep{burgarella_star_2005, noll_analysis_2009, boquien_cigale_2019} based on corrected aperture photometry performed in the five HST filters \citep{deger22}. 

To help break the age-reddening degeneracy, each cluster is classified as a `young nebular object' (YNO), part of the general cluster population, or an `old globular cluster' (OGC). This assignment is made based on the cluster's association with local H$\alpha$ emission from HST or ground-based observations and its location in the colour-colour diagram. Different sets of SED templates are applied to the photometry based on this decision tree classification. The age and $E(B-V)$ are determined as the values with the maximal posterior likelihood. 

Based on the derived age and $E(B-V)$, the cluster mass is calculated using the corresponding mass-to-light ratio. For age determination, the SED templates adopted include 10 linear age steps of 1~Myr for an age range of 1--10~Myr, and then 100 logarithmic steps of 0.03 dex for an age range of 10~Myr -- 13.75 Gyr. We refer readers to \citet{thilker_phangs-hst_2025} for a detailed description of the SED fitting procedure. The decision tree assignment leverages the HST H$\alpha$ observations \citep{chandar_2025} available at the time of publication (16 galaxies) and ground-based H$\alpha$ observations \citep{razza_2026} otherwise (20 galaxies). The lower spatial resolution of the ground-based H$\alpha$ imaging limits our ability to associate H$\alpha$ emission with individual clusters compared to the HST H$\alpha$ imaging. We assess the impact of this difference in Appendix \ref{sec:ground_vs_hst}.

We mask out star clusters in the central region of each galaxy, which we define using the environmental mask from \citet{querejeta_stellar_2021}. These are more severely affected by dust extinction. As a result, the completeness of the cluster catalogues is lower in these regions and the derived properties of the cluster are more uncertain \citep{whitmore_star_2021, turner_phangs-hst_2021}. Star-forming galaxy centres also tend to show a higher space density and a smaller separation between CO peaks, star forming regions, and star clusters \citep[e.g.][]{schinnerer_2023,sun_hidden_2024,tress_dynamical_2025,he_structure_2026}. As a result, the limited resolution of our CO maps affects analysis of these regions more severely.
We summarize the properties of the clusters selected and excluded in our analyses in Table \ref{tab:sample}. 

The completeness of the cluster catalogues is a combined function of stellar age and mass. In Fig. \ref{fig:mass_vs_age}, we show the cluster mass versus age diagram. The left panel shows the entire sample considered in this analysis while the right panel shows the young cluster population with age $\leq$10~Myr. 
The dotted line indicates a constant mass of 10$^{3.5}$ \solarmass that we later adopt as a mass threshold, above which we expect to recover a complete cluster sample across the age range of 1-- 10 Myr. In the right panel, we also separate young clusters into YNO clusters and non-YNO clusters. 

\subsection{Treatment of cluster ages in this paper}
\label{sec:agebins}

In the following analyses, we treat all YNO clusters as a single population rather than separating them according to their fitted ages of 1--5~Myr (see also Appendix \ref{sec:yno_age_test}). We divide the non-YNO clusters into age bins of 1--3, 4--5, 6--8, 9--10, 10--100, and $>100$~Myr. The boundaries of these bins are motivated by the positions of clusters along the BC03 model evolutionary track in the \textit{UVBI} colour-colour diagram. This treatment mitigates uncertainties in the fitted ages while allowing us to test whether the changing association with molecular gas across these age bins is consistent with an evolutionary sequence. Below, we describe the uncertainties in the fitted ages that motivate this treatment.

For YNO clusters, their identification as a whole is more robust than the age ordering within the 1--5~Myr interval. The \citet{thilker_phangs-hst_2025} SED fitting restricts the allowed ages to this range (right panel of Figure \ref{fig:mass_vs_age}), but distinguishing ages within it is sensitive to assumptions in the SED modelling. In particular, assumptions about the nebular contribution to the SED, including the H$\alpha$ escape fraction ($f_{\rm esc}$) and the radiation field, can affect the derived ages. Their adopted $f_{\rm esc}=0.5$ favours YNO ages of $\sim3\pm1$~Myr, with the precise ages shifting depending on the assumed nebular parameters. The true $f_{\rm esc}$ may be higher \citep{scheuermann_stellar_2026} or age-dependent.

There are additional indications that the age ordering within the YNO population is uncertain. A large fraction of YNO clusters assigned ages of 4--5~Myr occupy the region of the observed \textit{UVBI} colour-colour diagram associated with globular clusters before reddening correction. The extinction corrections from SED fitting are particularly uncertain for these highly reddened objects (Henny et al. in prep.). In fact, we find that YNO clusters assigned ages of 4--5~Myr are associated with higher CO intensities than those assigned ages of 1--3~Myr (Appendix \ref{sec:yno_age_test}), opposite to the expected progression if the fitted ages trace the dispersal of molecular gas with cluster age. We therefore treat the YNOs as a single young population, and expect YNO clusters to represent some of the youngest optical clusters in our sample.

For non-YNO clusters, the fitted ages provide useful statistical information, but the age assignments of individual young clusters are also uncertain. The non-nebular BC03 models evolve only weakly in \textit{UVBI} colour-colour space at the youngest ages, making it difficult to distinguish clusters with ages of 1--3~Myr. In particular, there is an excess of clusters assigned an age of 1~Myr because objects lying off the model evolutionary track can preferentially be assigned to the 1~Myr model, which has a slightly bluer $U-B$ colour \citep{whitmore_using_2011,hannon_h_2019,whitmore_empirical_2025}. We therefore combine clusters with fitted ages of 1--3~Myr into a single age bin.

The right panel of Figure \ref{fig:mass_vs_age} also shows a substantial population of massive non-YNO clusters with fitted ages $\leq3$~Myr. At these masses, stochastic sampling of the upper stellar initial mass function is unlikely to fully explain the absence of locally associated H$\alpha$ emission. Some of these clusters may be older than their fitted SED ages, while others may be genuinely young clusters for which the ionised gas has already been dispersed or is no longer sufficiently localized to satisfy the YNO selection. These uncertainties limit the precision of individual age estimates at the youngest ages, but the age bins can still provide a statistical clock for tracing cluster evolution.

The CO measurements provide an independent physical test of this statistical age sequence. If the age bins trace cluster evolution, we expect the amount of associated molecular gas to decrease toward older bins as clusters become increasingly dis-associated from their natal gas. We therefore examine how CO emission changes with age and cluster mass, while accounting for differences in the cluster mass distribution across the age bins.

\subsection{ALMA CO maps}

We use the ``flat'' version of the PHANGS--ALMA CO~(2-1) integrated intensity maps at 150 pc resolution, which adopt a spatially uniform noise treatment. We also perform similar analyses for a subset of seven galaxies that have CO(2-1) maps at 60 pc resolution for the resolution comparison in Appendix \ref{sec:co_res}. 
The ``flat'' maps are integrated intensity maps constructed by integrating with a mask that consists of a fixed velocity window around the large scale velocity field of the galaxy joined with any significant emission detected outside this window. Their production adopts the same principle as spectral stacking, which has been tested in PHANGS--ALMA by \citet{neumann_stacking_2023} and the ``flat'' maps have previously been used for comparison of ALMA and JWST data in \citet[][]{leroy_phangs_2023,chown_polycyclic_2025}. Compared to the other CO~(2-1) integrated intensity maps, these flat maps have approximately uniform noise, provide a measured value everywhere, and are designed to recover the total CO emission without introducing a selection bias against faint emission. To achieve this, they have higher noise than the more strictly masked maps from the PHANGS--ALMA pipeline. We expect this higher noise to average out when combining many measurements. The flat maps are ideal for statistical analysis that averages large populations of clusters.


\section{CO intensity measurements around star clusters \label{sec:profiles}}

\subsection{Age dependence of CO content in young clusters}
\label{subsec:Ico_hist_age}

\begin{figure*}
    \centering
    \includegraphics[width=0.45\textwidth]{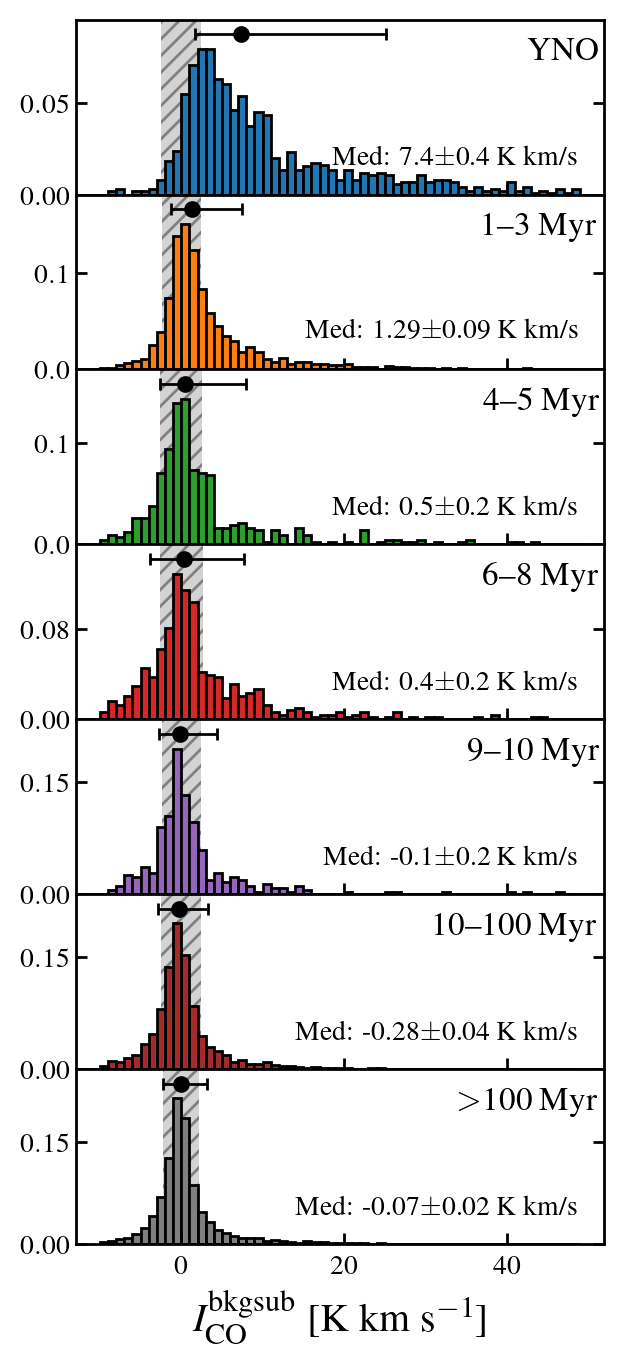}
    \includegraphics[width=0.47\textwidth]{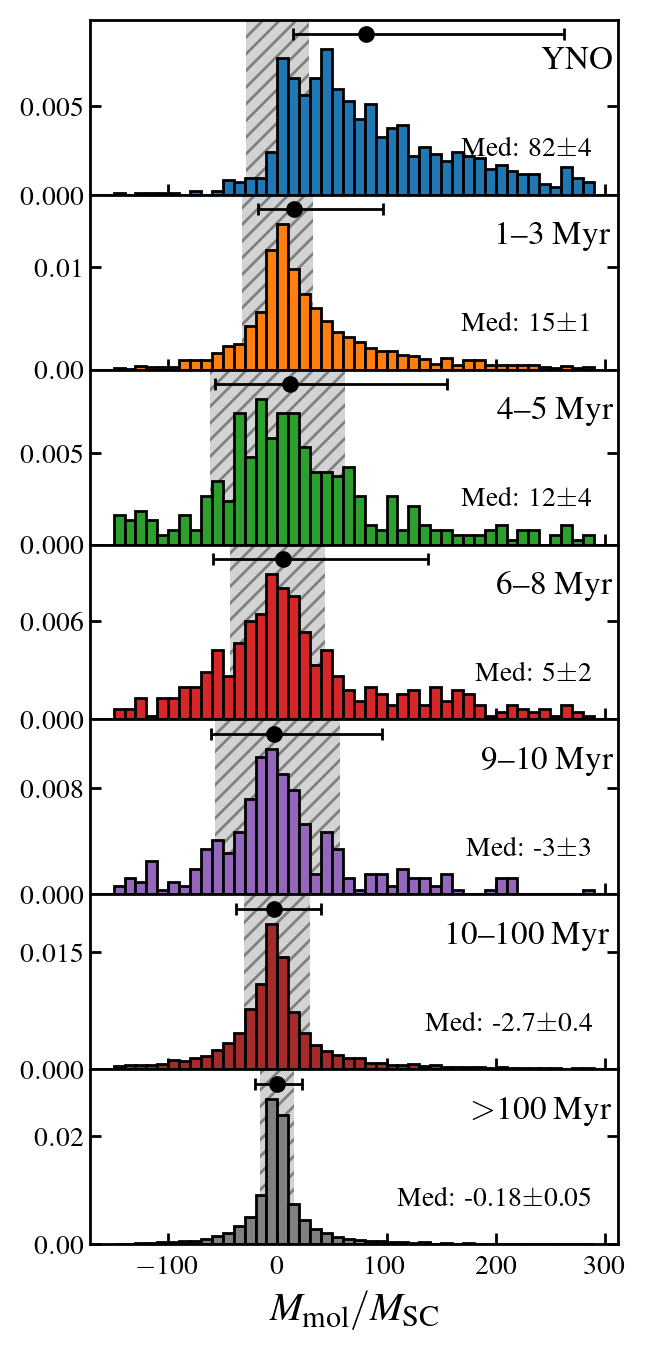}
    \caption{(\textit{Left}) The 1D histogram of background-subtracted CO intensity ($I_{\mathrm{CO}}^{\rm bkgsub}$) for clusters of different ages with mass above 10$^{3.5}$ \solarmass. The horizontal error bar indicates the median value and 16th-84th percentile range. The first age group only includes YNO clusters. The hatched region indicates the range of $I_{\rm CO}^{\rm bkgsub}$ values associated with $S/N < 3$ for individual CO measurements (this varies some from galaxy-to-galaxy). The lower right of each panel lists the median value and its uncertainty derived from bootstrapping and including both the sampling and measurement uncertainty. (\textit{Right}) Distribution of derived $M_{\rm mol}/M_{\rm SC}$ for the corresponding cluster age groups. The hatched regions again indicate the typical range for $S/N < 3$ in an individual CO measurement for a cluster with the median mass of the age group. Both panels show a clear evolutionary trend that indicates older clusters gradually dissociate from molecular gas. }
    \label{fig:cluster_co_intensity}
\end{figure*}

We record CO intensity $I_{\rm CO}$ at each cluster position as a measure of the molecular gas reservoir associated with these clusters. To isolate the ambient background CO emission from large-scale diffuse gas component (see \S \ref{subsec:profile_meas} and \S \ref{subsec:profile_outer}), we calculate the difference between $I_{\rm CO}$ and the median background level, 
\begin{equation}
I_{\rm CO}^{\rm bkgsub} (r=0) = I_{\rm CO}\left(r=0\right) - I^{\rm bkg}_{\rm CO}~,
\end{equation}
where $I_{\rm CO}^{\rm bkg}$ is the median $I_{\rm CO}$ within the deprojected annulus $r = 500{-} 1000$ pc from the cluster centre. 


Though we work in observational units, we note that for a fixed $150$~pc (FWHM) beam, cluster centre $I^{\rm bkgsub}_{\rm CO}$ can be translated to a molecular gas mass within the beam of
\begin{equation}
\label{eq:h2mass}
M_{\rm mol} = 1.7 \times 10^5~{\rm M_\odot} \left( \frac{I^{\rm bkgsub}_{\rm CO}}{\rm K~km~s^{-1}} \right) \left( \frac{\alpha_{\rm CO}^{2-1}}{\alpha_{\rm CO}^{2-1} ({\rm Milky~Way})} \right)~
\end{equation}
\noindent where we have assumed that neither $I_{\rm CO}^{\rm bksub}$ nor the beam area are corrected for inclination. Here $\alpha_{\rm CO}^{2-1}$ is the CO-to-H$_2$ conversion factor and $\alpha_{\rm CO}^{2-1} ({\rm Milky~Way}) = 6.7$ M$_\odot$ pc$^{-2}$ (K km s$^{-1}$)$^{-1}$ is a typical Milky Way conversion factor \citep[with fiducial $\alpha_{\rm CO}^{1-0}=4.35$ M$_\odot$ pc$^{-2} $ (K km s$^{-1}$)$^{-1}$ and CO(2-1)/(1-0) line ratio of 0.65; ][]{bolatto_co--h_2013}. In our analyses, we adopt a constant Milky-Way $\alpha_{\rm CO}^{2-1}$. We note that most $\alpha_{\rm CO}^{2-1}$ variation happens in the galaxy centre \citep[e.g.][]{schinnerer_molecular_2024, teng_star_2024}, which we exclude in our analyses. Therefore, we would not expect a large change in our result by applying varying $\alpha_{\rm CO}^{2-1}$. The impact of varying $\alpha_{\rm CO}^{2-1}$ will be explored in future work.

In Fig. \ref{fig:cluster_co_intensity} we show the histograms of $I_{\rm CO}^{\rm bkgsub}$ and derived $M_{\rm mol}/M_{\rm SC}$ for each age bin and restricting to $M_{\rm SC} \geq 10^{3.5}$~M$_\odot$, where we expect the cluster catalogues to be complete for young clusters (see Figure~\ref{fig:mass_vs_age}). Table \ref{tab:Idiff_results} summarizes these results.

Young clusters associated with H$\alpha$ (YNOs) show the highest median peak $I_{\rm CO}^{\rm bksub}$=7.4 K~km~s$^{-1}$ and median ratio of $I_{\rm CO}^{\rm bksub}/M_{\rm SC}=4.8\times10^{-4}$ K~km~s$^{-1}$~M$_{\odot}^{-1}$. In physical terms, these values correspond to molecular gas mass ($M_{\rm mol}$) of $1.3\times 10^{6}$ \solarmass and gas-to-cluster mass ratio ($M_{\rm mol}/M_{\rm SC}$) of $82$, i.e., that the cluster mass is $\approx 1.2\%$ of the spatially associated molecular gas. Both the absolute and normalized CO content drop, to $2.2\times10^5$ \solarmass and $15$, when considering non-YNO clusters of 1-3 Myr old and then further down to $8.5\times10^4$ \solarmass and $12$ when considering 4{-}5 Myr old clusters. This suggests that the YNOs are the clusters least processed by stellar feedback, and thus perhaps also the youngest clusters.
However, because the YNO classification requires locally associated H$\alpha$ emission, there must be some gas present for these objects by construction. Our measurement on its own does not distinguish how much the enhanced CO associated with YNOs reflects this selection effect as opposed to truly catching the youngest regions.

In comparison, 1-3 Myr old clusters in \citet{thilker_phangs-hst_2025} without associated H$\alpha$ (i.e. they are not YNOs) have $5.6$ times lower $I_{\rm CO}^{\rm bksub}$ and $5.5$ times lower $M_{\rm mol}/M_{\rm SC}$ than the YNOs, on average. If the YNOs do represent precursors to the non-YNO young clusters, this is additional evidence for significant gas clearing within the first $\sim 3$~Myr. Such a short timescale for gas clearing suggests that pre-supernovae (pre-SNe) feedback (e.g.\ stellar winds, photoionisation and radiation pressure) plays an important role in gas dispersal, since the first supernova is expected to explode at $\sim$3 Myr \citep[for a 100~\solarmass star,][]{zapartas_delay-time_2017}. This is consistent with various theoretical and simulation works \citep{inutsuka_formation_2015, peters_silcc_2017, grudic_when_2018, jeffreson_momentum_2021, guszejnov_effects_2022, farias_stellar_2024, tress_rhea_2026} that highlights the dominant and universal role of pre-SNe feedback in dispersing gas for clusters of different masses. Observational studies of H~II regions \citep[e.g.][]{barnes_comparing_2021, olivier_evolution_2021,pathak_linking_2025} also suggest that the pressure induced by this pre-SNe feedback can effectively expand and clear the surrounding interstellar medium. 

Clusters with age 4--5 Myr have a similar $M_{\rm mol}/M_{\rm SC}$ as 1--3 Myr old clusters, though the absolute value of $I_{\rm CO}^{\rm bkgsub}$ (and absolute amount of molecular gas $M_{\rm mol}$) is lower for the 4--5 Myr clusters. This reflects that clusters at 4--5 Myr have lower mass compared to clusters at 1--3 Myr because of the lower mass to light ratio for older clusters and approximately fixed magnitude limit (see Fig. \ref{fig:mass_vs_age}). The similar $M_{\rm mol}/M_{\rm SC}$ ratio at 4--5 Myr and 1--3 Myr suggests that a certain amount of molecular gas remains attached to the clusters through this period, and may be further cleared by later supernova explosions. 

The situation at $6{-}8$~Myr is ambiguous. Though the signal is weak, these clusters do show higher median CO intensity and a larger fraction of clusters with high $M_{\rm mol}/M_{\rm SC}$ ratio ($f_{\rm high}=$20\% for clusters with $M_{\rm mol}/M_{\rm SC} \geq 100$, Table \ref{tab:Idiff_results}) compared to 9--10 Myr clusters ($f_{\rm high}=14\%$ for the same condition). For massive clusters ($M_{\rm SC} \geq 10^4$ \solarmass), we also find the signature of central gas depletion (Fig. \ref{fig:stack_Idiff_mass}) which is probably due to further clearing of gas by supernova explosion. 

After $9{-}10$~Myr both the median $I_{\rm CO}^{\rm bksub}$ and $I_{\rm CO}^{\rm bksub}/M_{\rm SC}$ show no gas associated with clusters, on average. This suggests that by $\sim 10$~Myr, older clusters have become spatially dissociated from their natal molecular cloud either due to stellar feedback or drift of the cluster relative to the cloud. In \S \ref{subsec:feedback_drift}, we argue that cluster drift alone can not reduce the cluster associated $I_{\rm CO}^{\rm bkgsub}$ and $M_{\rm mol}/M_{\rm SC}$ to the observed values. Therefore, our results suggest a complete clearing of gas after 9--10 Myr since cluster formation, on average.  

We note that our measurements in Fig. \ref{fig:cluster_co_intensity} do not take into account the uncertainty in cluster ages. In Appendix \ref{sec:age_uncertainty} we perform Monte-Carlo tests and find that our observed trend appears robust against these age uncertainties.

\subsection{Mass dependence of CO content in young and old clusters}
\label{subsec:Ico_vs_mass}

\begin{figure*}
\centering
\includegraphics[width=0.9\textwidth]{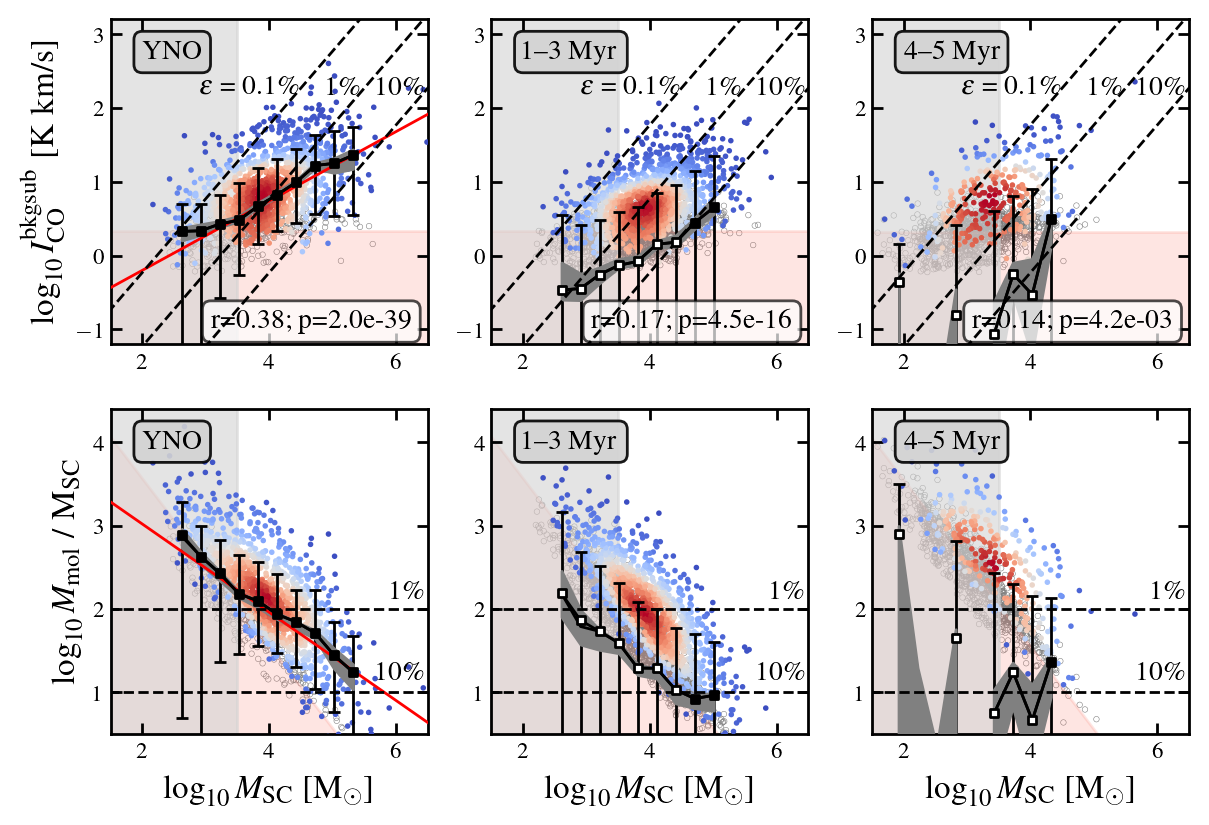}
\caption{(\textit{Top}) The background subtracted $I_{\rm CO}$ versus the cluster mass. The coloured points show significant CO detections (S/N$\geq 3$) colour coded by the number density of points. The empty gray circles indicate the upper limit of non-significant CO detections. The salmon shade indicates the region where the typical $S/N$ of the CO~(2-1) data is $< 3$. The light-gray shaded region indicates the mass range below our threshold of 10$^{3.5} \: {\rm M_{\odot}}$, where we consider our cluster population recovery incomplete for clusters younger than 10 Myr.  
The dashed lines indicate a constant $I_{\rm CO}^{\rm bkgsub}$/$M_{\rm mol}$ ratio and $\epsilon$ represents the corresponding observed cluster formation efficiency (Eq. \ref{eq:efficiency}). The error bars indicate the binned median with 16th--84th percentile range and the dark gray shaded region indicates the uncertainty of the median. The filled and open squares indicate mass bins with CO detection fraction above and below 50\%, respectively. The red lines indicate the \texttt{linmix} power-law fit to the observed trend (\S \ref{subsec:Ico_vs_mass}). 
(\textit{Bottom}) The derived gas mass to cluster mass ratio versus cluster mass. All symbols and shaded regions have the same meaning as in the top panel. }
\label{fig:Idiff_mass_young}
\end{figure*}

\begin{figure*}
\centering
\includegraphics[width=0.9\textwidth]{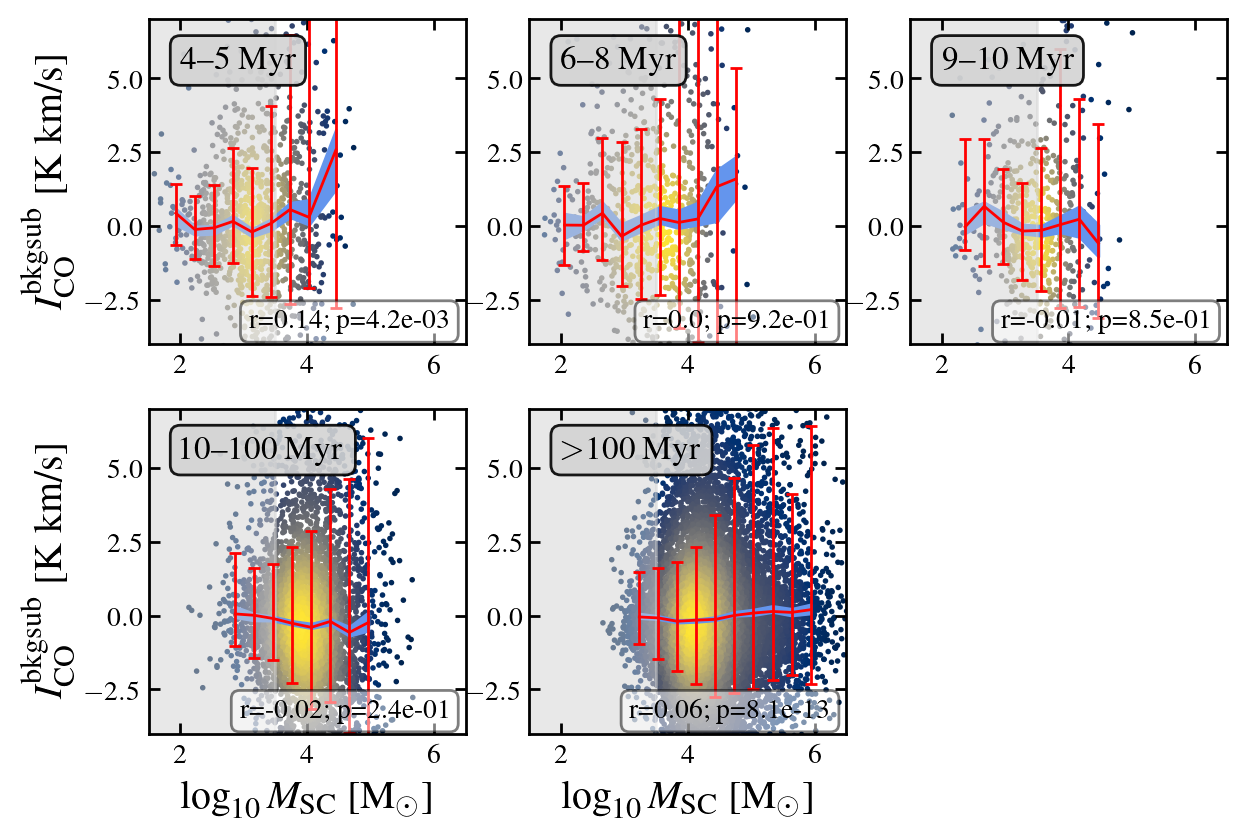}
\caption{The background subtracted $I_{\rm CO}$ versus the cluster mass for the $\geq$4 Myr old cluster population. The points are colour coded by the number density. 
The red error bars shows the median and 16th-84th scatter. The blue shaded region indicates the median uncertainty for each bin. The light-gray shaded region indicates mass range below our common threshold of 10$^{3.5}$ \solarmass. For clusters older than 6 Myr, we see that $I_{\rm CO}^{\rm bkgsub}$ has little mass dependence and has median values close to zero. }
\label{fig:Idiff_mass_old}
\end{figure*}

In Fig. \ref{fig:Idiff_mass_young} we show the correlation between $I_{\rm CO}^{\rm bkgsub}$ and cluster mass for young clusters in different age groups. In Fig. \ref{fig:Idiff_mass_old} we show the same for the older clusters. We report the Spearman correlation coefficient and p-value in Table \ref{tab:Idiff_results}.  

The figures and table show that clusters younger than 6 Myr exhibit significant positive correlations between cluster mass, $M_{\rm SC}$, 
and $I_{\rm CO}^{\rm bkgsub}$. YNO clusters show the strongest correlation (Spearman $r$ of 0.38). 
Then the correlation gets weaker and both the absolute and normalized $I_{\rm CO}^{\rm bkgsub}$ decline for non-YNO clusters of 1--3 Myr and 4--5 Myr. For clusters older than 6 Myr, $I_{\rm CO}^{\rm bkgsub}$ shows no significant mass dependence with a median value of approximately 0~K~km~s$^{-1}$. 


Dashed lines in Fig. \ref{fig:Idiff_mass_young} show fixed ratios of $I_{\rm CO}^{\rm bksub}$ to cluster mass. The binned trends for YNO clusters do not follow these lines, indicating a changing ratio of gas to star mass as a function of cluster mass. We also perform a \textsc{linmix} \citep{kelly2007} fit to the $I_{\rm CO}^{\rm bksub}$ vs $M_{\rm SC}$ relation (including 3$\sigma$ upper limits) for YNO clusters (red solid line) and obtain 
\begin{equation}
\log_{10} I_{\rm CO}^{\rm bkgsub}[{\rm K\ km\ s^{-1}}] = 0.48 \times \log_{10} M_{\rm SC}[{\rm M_\odot}] - 1.14
\label{eq:Idiff_vs_mass}
\end{equation}
The sub-linear power-law slope indicates decreasing $I_{\rm CO}^{\rm bksub}$/$M_{\rm SC}$ ratio as cluster mass increases. Converting $I_{\rm CO}^{\rm bksub}$ to molecular gas mass $M_{\rm mol}$ (Eq. \ref{eq:h2mass}), we obtain the cluster mass dependence of gas-to-cluster mass ratio 
\begin{equation}
\label{eq:yno_eff}
\log_{10}\left(M_{\rm mol}/M_{\rm SC}\right) = -0.52 \times \log_{10} M_{\rm SC}[{\rm M_\odot}] + 4.1
\end{equation}
For non-YNO clusters of 1--3 Myr, we see a similar cluster mass dependence of $M_{\rm mol}/M_{\rm SC}$ from the binned trend (lower-middle panel of Fig. \ref{fig:Idiff_mass_young}). However, all the mass bins below 10$^4$ \solarmass have CO detection fraction below 50\% (denoted as open square). While this does not formally render the stacked measurements invalid, we worry that the measured trend could be influenced by the lower individual $S/N$ and do not report a fit. 

If clusters do not drift from their birth sites (see \S \ref{subsec:feedback_drift}), then this ratio reflects the combination of the efficiency with which clusters form from molecular gas ($M_{\rm SC} \sim \epsilon_{\rm form} M_{\rm mol}^{\rm init}$) and the efficiency with which stellar feedback disperses that gas over the age of the cluster ($M_{\rm mol}^{\rm current} \sim (1 - \epsilon_{\rm feedback} (\tau)) M_{\rm mol}^{\rm init}$). Our observed cluster formation efficiency can be expressed as
\begin{equation}
\epsilon_{\rm obs} = \left(\frac{M_{\rm mol}}{M_{\rm SC}}\right)^{-1} = \frac{\epsilon_{\rm form}}{(1 - \epsilon_{\rm feedback})}
\label{eq:efficiency}
\end{equation}
We note that this efficiency also does not account for double-counting of the star clusters within the 150~pc CO~(2-1) beam in the calculation. We discuss this caveat and its possible implication in \S \ref{subsec:formation_efficiency}. 

If stellar feedback has not been effective yet for YNO clusters, then $\epsilon_{\rm obs}$ reflects the formation efficiency. Then the sublinear slope of the fit in Equation \ref{eq:Idiff_vs_mass} suggests that $\epsilon_{\rm form}$ becomes higher for more massive clusters. This is consistent with theoretical expectation, which predicts an increase in both the star formation efficiency from gas to stars ($\epsilon_{\rm SFE}$) and the bound fraction of formed stars within star clusters ($\Gamma$) with higher GMC surface density \citep[e.g.][]{adams_theoretical_2000, fall_stellar_2010, kruijssen_fraction_2012, li_disruption_2019, grudic_when_2018, grudic_model_2021}. 

Meanwhile for non-YNO clusters aged 1--5 Myr, Figure \ref{fig:Idiff_mass_young} shows higher $\epsilon_{\rm obs}$ ($\sim$10\%) compared to that of YNO clusters of the same mass. The most natural explanation for this is an increase in $\epsilon_{\rm feedback}$ over time, as stellar feedback disperses the natal gas from which the clusters formed.

\subsection{Galaxy-by-galaxy variation in observed YNO cluster formation efficiency}
\label{subsec:gal_by_gal}

\begin{figure*}
\centering
\includegraphics[width=0.35\textwidth]{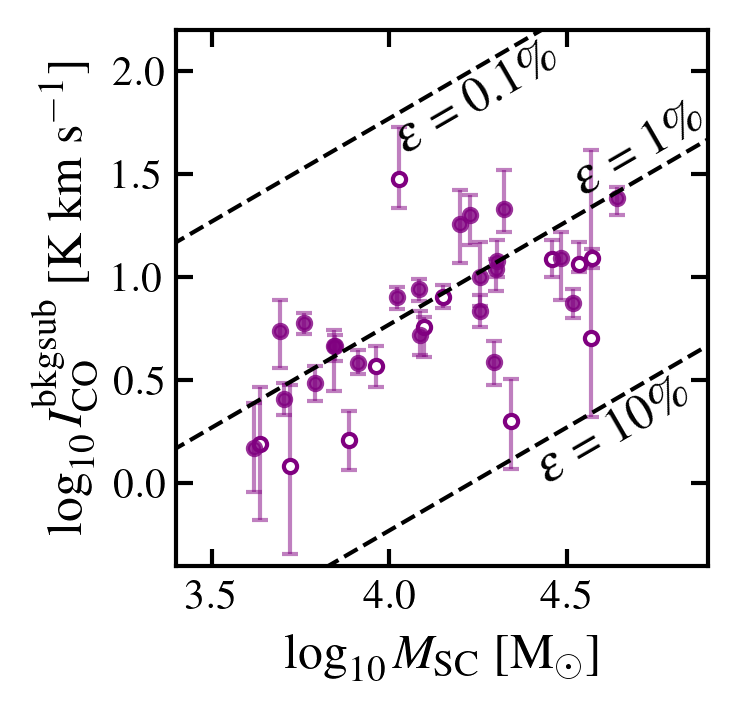}
\includegraphics[width=0.6\textwidth]{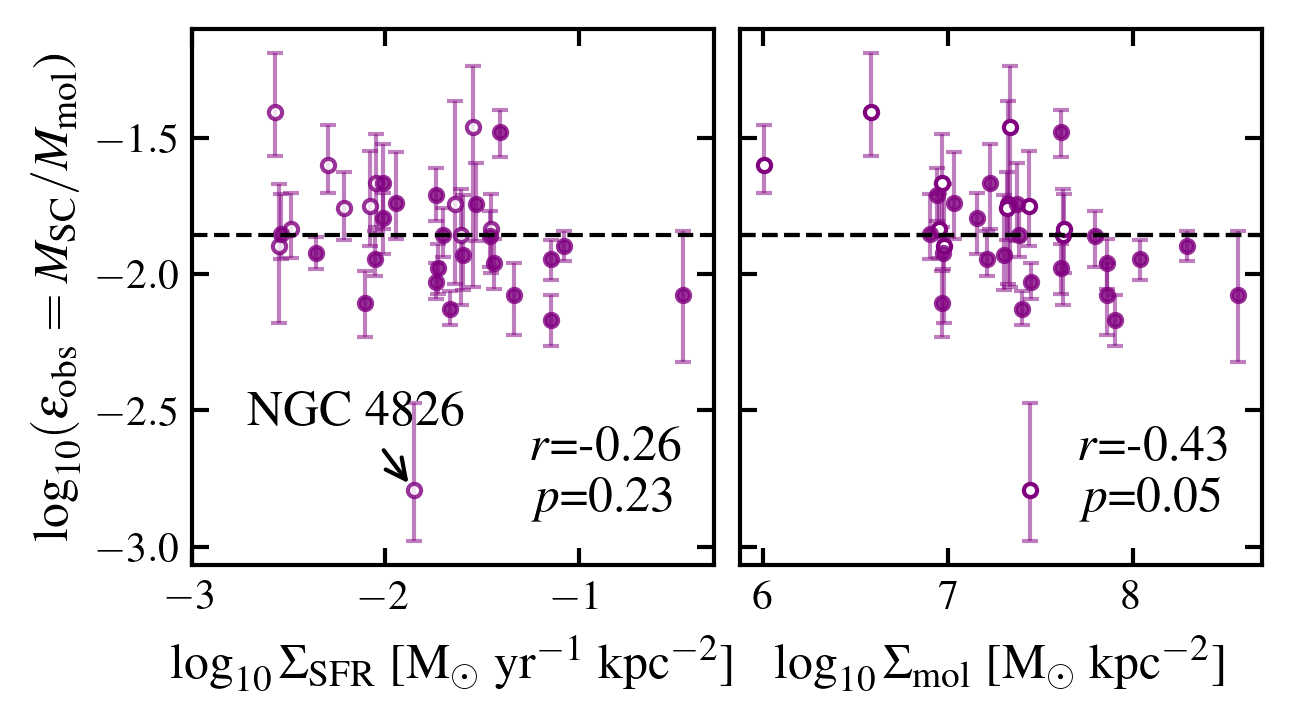}
\caption{(\textit{Left}) The median $I_{\rm CO}^{\rm bkgsub}$ versus median cluster mass for YNO clusters of individual galaxies. Error bars indicate the uncertainty on the median. Empty circles indicate galaxies with less than YNO clusters. 
The dashed lines indicate constant cluster mass to gas mass ratio (Eq. \ref{eq:efficiency}). (\textit{Middle and right}) The observed efficiency from gas to clusters ($\epsilon_{\rm obs} \equiv M_{\rm SC}/M_{\rm mol}$, Eq.\ref{eq:efficiency}) as a function of the SFR surface density ($\Sigma_{\rm SFR}$) and molecular gas surface density ($\Sigma_{\rm mol}$) drawn from \citet{leroy_phangs-alma_2021}. The dashed line indicates the median $\epsilon_{\rm obs}$ ($\sim$1.4\%) for all galaxies. In the lower-right corner we report the Spearman correlation coefficient ($r$) and $p$-values ($p$) for galaxies with $>10$ YNO clusters. }
\label{fig:gal_indvd}
\end{figure*}


So far, our analyses combine measurements from all galaxies in our sample. But our targets vary in mass, intensity of star formation, dynamical environment, and structure in ways that might impact the cluster formation process. Therefore, we calculate the median $I_{\rm CO}^{\rm bkgsub}$ and $M_{\rm mol}/M_{\rm SC}$ (i.e. $1/\epsilon_{\rm obs}$) for YNO clusters ($M_{\rm SC} \geq 10^{3.5}$ \solarmass) for each individual galaxy. In the left panel of Fig. \ref{fig:gal_indvd}, we correlate these individual-galaxy $I_{\rm CO}^{\rm bkgsub}$ and $M_{\rm SC}$, and we further summarize star cluster properties and related CO measurements for individual galaxies in Appendix \ref{sec:gal_indvd}.

In Fig. \ref{fig:gal_indvd}, the median of galaxy-averaged $\epsilon_{\rm obs}$ is $\sim$1.4\%, which is very close to the overall median of 1.2\% (Table \ref{tab:Idiff_results}). For galaxies with more than 10 YNO clusters, we calculate the 16th-84th scatter of $\epsilon_{\rm obs}$ within individual galaxies, as well as the scatter among galaxy-averaged $\epsilon_{\rm obs}$. We find the typical 1$\sigma$ scatter within individual galaxies is $\sim$0.53 dex (median value of galaxies), which is larger by a factor of three than the galaxy-by-galaxy scatter of $\sim$0.17 dex. This suggests that the overall scatter of the $I_{\rm CO}^{\rm bkgsub}$ versus $M_{\rm SC}$ correlation is mainly driven by the internal variation within galaxies.  

In the right panels of Fig. \ref{fig:gal_indvd}, we plot $\epsilon_{\rm obs}$ as a function of galaxy-averaged $\Sigma_{\rm SFR}$ and $\Sigma_{\rm mol}$ \citep[from ][]{leroy_phangs-alma_2021}. We find a slight anti-correlation between $\epsilon_{\rm obs}$ and both quantities, though both $p$-values are large and $\epsilon_{\rm obs} \propto \left(I_{\rm CO}^{\rm bkgsub}\right)^{-1}$ may be indirectly anti-correlated with global $\Sigma_{\rm mol}$ by construction. The weakness of these relations suggests that $\epsilon_{\rm obs}$ is relatively independent of global galaxy properties. 

Figure \ref{fig:gal_indvd} shows one outlier galaxy with extremely low $\epsilon_{\rm obs}$. This is NGC 4826 (the ``Evil Eye Galaxy'', M64), which has undergone a recent merging event \citep{braun_counter_1992} that creates a counter-rotating molecular disk within the central 1~kpc region. While the molecular surface density is high, the gas distribution is smooth. This is probably due to the strong shear caused by rotation \citep[e.g. as in][]{liu_wisdom_2021}. This dynamical stabilization may keep the gas from collapsing and forming star clusters. As NGC 4826 is very nearby, we highlight this target as an ideal case to follow up and test theories related to dynamical suppression of cluster formation.

\subsection{Stacked intensity profiles show CO associated with young clusters}
\label{subsec:profile_meas}

\begin{figure*}
    \centering
    \includegraphics[width=0.9\textwidth]{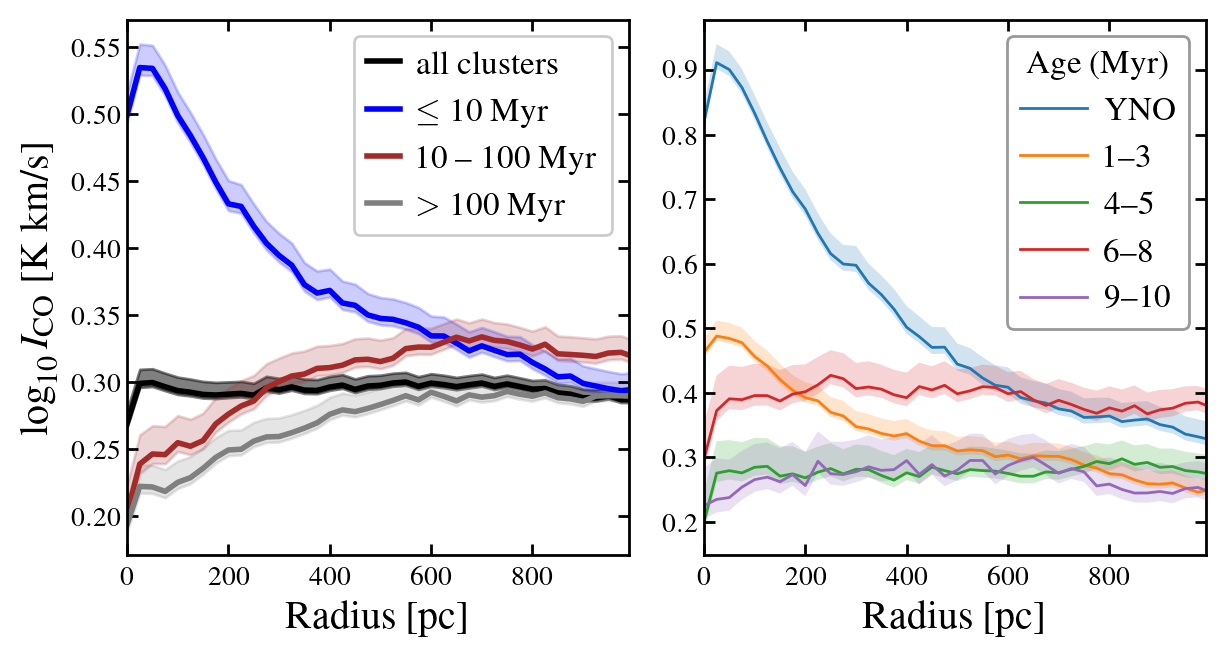}
    \caption{(\textit{Left}) Stacked median CO~(2-1) intensity profiles around the star clusters of three age groups: younger than 10~Myr (\textit{blue}), between 10 and 100~Myr (\textit{brown}) and older than 100~Myr (\textit{gray}). The shaded region indicates the uncertainty of the median derived from propagating the uncertainties in the original CO~(2-1) maps. (\textit{Right}) The stacked intensity profiles for star clusters younger than 10~Myr are divided into finer age groups as discussed in Section~\ref{sec:agebins}.}
    \label{fig:stacked_profile}
\end{figure*}

\begin{table*}[htb]
\caption{\label{tab:Idiff_results} Cluster properties of different age groups ($M_{\rm SC} \geq 10^{3.5}$\solarmass)}
\centering
\renewcommand{\arraystretch}{1.4}
\setlength{\tabcolsep}{3pt}
\begin{tabularx}{\textwidth}{ccccccccccccc}
\hline\hline 
Age & $N_{\mathrm{SC}}$& \multicolumn{3}{c}{$\log_{10} M_{\rm SC}$  } & \multicolumn{3}{c}{$I^{\rm bkgsub}_{\rm CO}$} & \multicolumn{3}{c}{$M_{\rm mol}$ / $M_{\rm SC}$} & ($r$, $p$) & $f_{\rm high}$ \\
$[\rm Myr]$ & & \multicolumn{3}{c}{[\solarmass]} & \multicolumn{3}{c}{[K km s$^{-1}$]} & \multicolumn{3}{c}{} &  & $(M_{\rm mol}/M_{\rm SC}\geq 100)$ \\
 & &  Med & 16th & 84th & Med & 16th & 84th & Med & 16th & 84th &   \\
(1) & (2) & (3) & (4) & (5) & (6) & (7) & (8) & (9) & (10) & (11) & (12) & (13) \\
\hline
YNO & 1113 & 4.1 & 3.7 & 4.7 & $7.4\pm 0.3$ & 1.7 & 25.2 & $82\pm 4$ & 15 & 263 & (0.38, 2e-39) & 0.43 \\
1--3 & 2359 & 4.1 & 3.7 & 4.5 & $1.29\pm 0.09$ & -1.2 & 7.5 & $15\pm 1$ & -18 & 96 & (0.17, 4e-16) & 0.15 \\
4--5 & 439 & 3.8 & 3.6 & 4.2 & $0.5\pm 0.2$ & -2.6 & 7.9 & $12\pm 4$ & -57 & 156 & (0.14, 4e-03) & 0.23 \\
6--8 & 499 & 4.0 & 3.7 & 4.4 & $0.4\pm 0.2$ & -3.9 & 7.7 & $5\pm 3$ & -59 & 138 & (0.0, 9e-01) & 0.21 \\
9--10 & 358 & 3.8 & 3.6 & 4.2 & $-0.1\pm 0.2$ & -2.8 & 4.4 & $-3\pm 3$ & -61 & 96 & (-0.01, 8e-01) & 0.15 \\
10--100 & 4422 & 4.1 & 3.7 & 4.6 & $-0.28\pm 0.04$ & -2.9 & 3.4 & $-2.7\pm 0.4$ & -38 & 40 & (-0.02, 2e-01) & 0.08 \\
$>$ 100 & 12672 & 4.4 & 3.9 & 5.1 & $-0.07\pm 0.02$ & -2.3 & 3.1 & $-0.18\pm 0.05$ & -20 & 22 & (0.06, 8e-13) & 0.04 \\
\hline
\end{tabularx}
\tablefoot{Columns: (1) Age range of the cluster group as discussed in Section~\ref{sec:agebins}. (2) The number of clusters in each age group. (3) (4) (5) The median and the 16--84\% range of the cluster mass in that group. (6) (7) (8) The median, uncertainty in the median, and 16--84\% range of the background subtracted $I^{\rm bkgsub}_{\rm CO}$ for clusters in that group. (9) (10) (11) The median, the uncertainty in the median, and 16--84\% range of the mass ratio between molecular gas mass derived from  $I^{\rm bkgsub}_{\rm CO}$ and the cluster mass from the catalogue for that group. (12) The Spearman coefficient and $p$-value describing the correlation between cluster mass and $I^{\rm bkgsub}_{\rm CO}$. (13) The fraction of $M_{\rm mol}/M_{\rm SC}$ ratios in that group with a high value, defined as $>100$. }
\end{table*}

We quantify the CO~(2-1) distribution around each population of clusters by stacking the radial profiles of CO. We define a series of elliptical annuli around each cluster position that take into account the inclination and position angle of the galaxy and have radial bin widths of 25 pc (compare to the 150~pc FWHM of the homogenized CO beam). 
We then measure the median $I_{\mathrm{CO}}$ within each annulus out to 1~kpc. 

In Fig.~\ref{fig:stacked_profile}, we show the median CO intensity profiles for clusters in different age groups. In the left panel, we divide the full sample into three groups -- clusters younger than 10~Myr, clusters with an age of 10 -- 100~Myr and clusters older than 100~Myr \citep[e.g. following][]{grasha_connecting_2018, grasha_spatial_2019, turner_phangs_2022}. Young clusters (age $<10$~Myr) have central gas concentrations and declining intensity with radius. In contrast, older clusters (age $>10$~Myr) show the opposite trend, with no systematic enhancement of gas at the peaks and some evidence of a depression, which we discuss below.
These results are consistent with previous two-point correlation studies \citep[e.g.,][]{grasha_connecting_2018, grasha_spatial_2019, turner_phangs_2022} that show GMC locations to be more strongly correlated with young clusters than old clusters. 

In the right panel of Fig.~\ref{fig:stacked_profile}, we divide clusters younger than 10~Myr into the age bins defined in Section~\ref{sec:agebins}. 
Clusters younger than 4~Myr have a strong central gas concentration, while clusters with ages 4{-}10 Myr show a flatter CO distribution that resembles the results for older $> 10$~Myr clusters (but see below). YNO clusters, i.e., those with associated H$\alpha$ and age 1--5~Myr, have the strongest peak value and strongest contrast with the large-scale background, indicating the strongest spatial association with CO intensity for this population.

\subsection{Stacked radial profiles and 2D images of background subtracted CO intensity}
\label{subsec:Idiff_profile}

\begin{figure*}
\centering
\includegraphics[width=\textwidth]{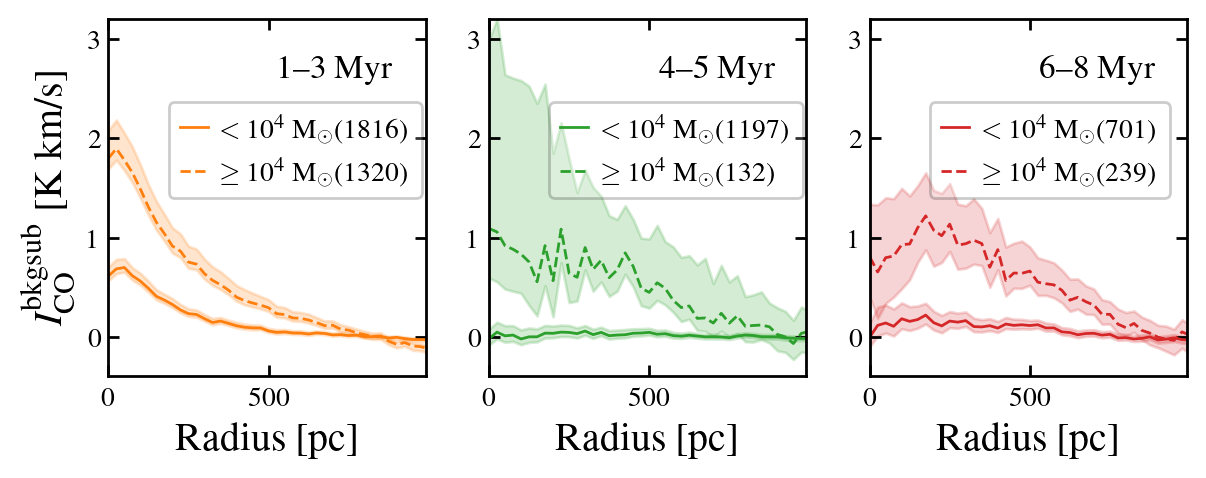}
\caption{The stacked median radial profiles of $I_{\rm CO}^{\rm bkgsub}$ for non-YNO clusters of age 1--8 Myr. The solid and dashed lines indicate the median profiles of clusters with mass lower and higher than 10$^4$ \solarmass. The shaded regions indicate the median uncertainty. Different colours indicate clusters of different age groups: (\textit{orange}) 1--3 Myr, (\textit{green}) 4--5 Myr and (\textit{red}) 6--8 Myr. The number in the bracket indicates the number of clusters of each sub-population. }
\label{fig:stack_Idiff_mass}
\end{figure*}

\begin{figure*}
\centering
\includegraphics[width=\textwidth]{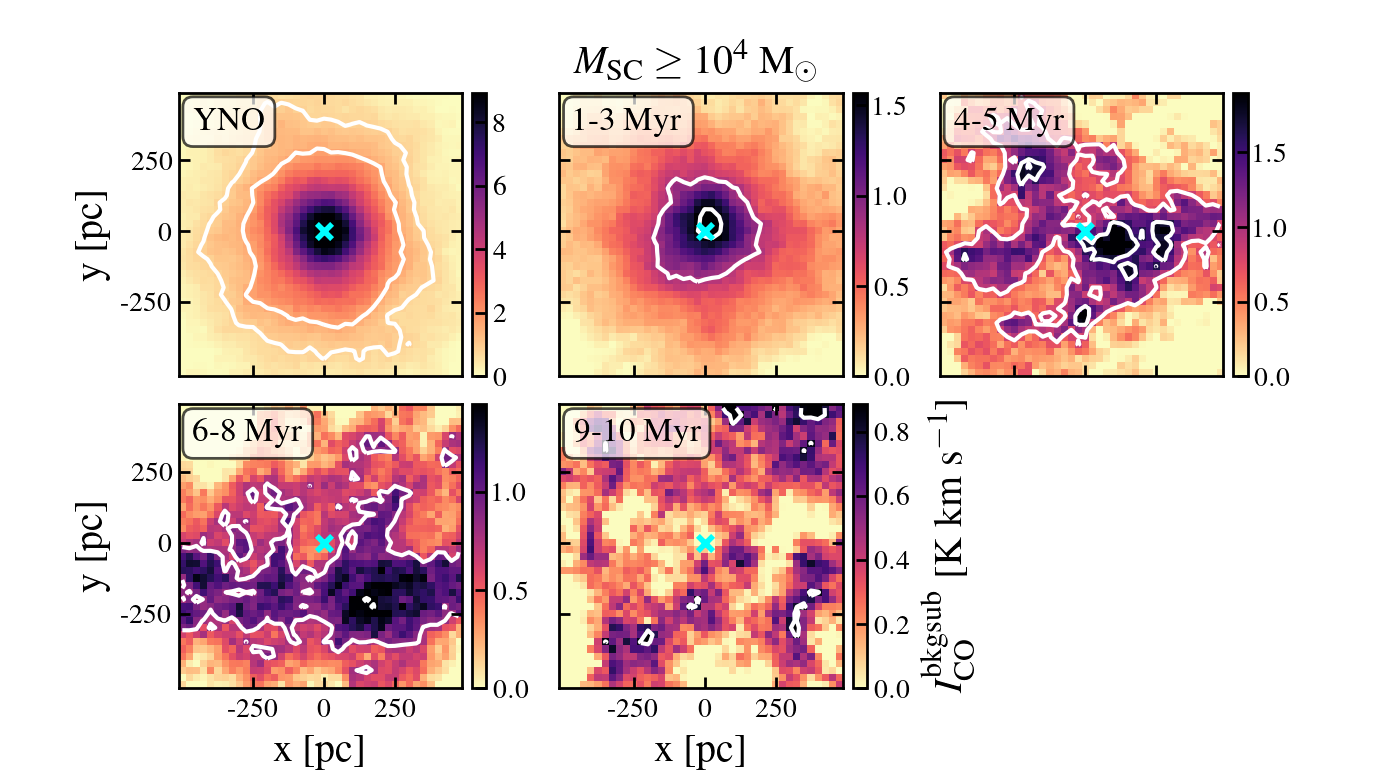}
\caption{The stacked median of the 1~kpc $\times$ 1~kpc cutout CO maps centred around individual clusters with mass $\geq 10^4$ \solarmass from different age groups. Background subtraction is applied for each cutout image before the stacking. 
The cyan cross indicate the cluster centre position. The white contours indicate the typical 1$\sigma$ and 2$\sigma$ noise levels (0.82 and 1.64 K~km~s$^{-1}$). }
\label{fig:image_stacking}
\end{figure*}


In Fig. \ref{fig:stack_Idiff_mass}, we show the stacked median of the radial profiles of $I_{\rm CO}^{\rm bkgsub} (r)$ for non-YNO clusters of different age and mass groups. For all age bins from 1 -- 8 Myr, we find higher $I_{\rm CO}^{\rm bkgsub} (r)$ for more massive clusters ($M_{\rm SC}\geq 10^4$ \solarmass). This can also be seen from Figs. \ref{fig:Idiff_mass_young} and \ref{fig:Idiff_mass_old}. These show increased $I_{\rm CO}^{\rm bkgsub}$ above $M_{\rm SC} \approx 10^4$~M$_\odot$ in all age bins $\leq 6{-}8$~Myr, though there are fewer clusters with such high $M_{\rm SC}$ in the 4--5 and 6--8 Myr bins. 

On the other hand, the same profiles and scatter plots show little or no $I_{\rm CO}^{\rm bkgsub}$ associated with lower mass clusters in the 4--5 and 6--8 Myr bins. In Fig. \ref{fig:stack_Idiff_mass} for $M_{\rm SC} < 10^4$~M$_\odot$ clusters, the stacked profile appears flat with $I_{\rm CO}^{\rm bkgsub} \approx 0$~K~km~s$^{-1}$ already at 4--5 Myr. In Figs. \ref{fig:Idiff_mass_young} and \ref{fig:Idiff_mass_old} the median central $I_{\rm CO}^{\rm bkgsub}$ is almost flat at $0$~K~km~s$^{-1}$ for $M_{\rm SC} < 10^4$~M$_\odot$ and ages 4--5 or 6--8 Myr. In these same age bins the (rarer) high mass clusters still show association with molecular gas concentrations. 

These measurements imply that massive clusters remain associated with cold gas longer than lower mass clusters. This could reflect a difference in their birth environments. We expect low-mass clusters to form from low-mass GMCs, which tend to live in isolated environments with low GMC number density \citep{he_structure_2026}. In this case, once stellar feedback clears the gas from the central birth GMC after $\sim$4 Myr, the intensity profile will become flat. In contrast, high-mass clusters form from massive GMCs that live in crowded environments and cluster with one another. In this case we expect nearby GMCs, not just the cloud from which the cluster forms, to contribute to the CO emission in our radial profile measurements. One would expect it to take longer for stellar feedback to clear a cluster of massive GMCs than a single cloud. We note that this difference could also be due to our CO sensitivity limit, which might fail to recover the central gas concentration for low-mass clusters. High-mass clusters ($M_{\rm SC} \geq 10^4$ \solarmass) have a median $M_{\rm mol}/M_{\rm SC}$ ratio of $11 \pm 5$, while low-mass clusters ($M_{\rm SC} < 10^4$ \solarmass) have the ratio of $2 \pm 5$, and hence the ratio difference we see is barely within the error margin. 

For massive clusters with ages 6--8 Myr, the $I_{\rm CO}^{\rm bkgsub}$ profile shows a central dip. This might indicate the central depletion of gas due to stellar feedback near the site of the cluster. To investigate this further, we perform direct image stacking of the CO maps for $M_{\rm SC}\geq 10^4$ \solarmass\ clusters. Fig. \ref{fig:image_stacking} shows the median of $1~{\rm kpc}\times{\rm 1~kpc}$ regions of the CO maps (with local background subtraction) centred on individual clusters from different age groups. 

In the image stack, YNO clusters and non-YNO clusters of 1--3 Myr show a strong and symmetric central gas concentration, which is consistent with our radial profile stacking results. In contrast, clusters at 4--5 Myr and 6--8 Myr old show a more extended and asymmetric CO emission distribution. This may indicate the role of stellar feedback in re-shaping the surrounding gas. Clusters at 9--10 Myr show a complete central gas depletion with $I_{\rm CO}^{\rm bkgsub}$ dropping to zero. The morphology change for massive clusters between 1--3 Myr and 4--5 Myr is interesting given that the two age groups have similar $M_{\rm mol}/M_{\rm SC}$ (Fig. \ref{fig:Idiff_mass_young}). This may indicate that stellar feedback redistributes the gas and makes it patchier before clearing the material, similar to what is seen for H$\alpha$ around star clusters \citep{hannon22}. Alternatively, since this image is a stack the patchiness may reflect heterogeneity among the different clusters being combined, indicating that over this time frame the impact of feedback and drift are highly variable.

We note that the 150~pc resolution of our CO~(2-1) data means that these profiles reflect the clustering of molecular clouds or the structure of giant molecular associations more than any sub-cloud structure \citep[see][]{he_structure_2026}. PHANGS--ALMA has a smaller subset of higher physical resolution maps, and in Appendix \ref{sec:co_res}, we compare the stacked $I_{\rm CO}^{\rm bkgsub}$ at 60 pc and 150 pc resolution for a subset of galaxies with both resolutions available. We find that $I_{\rm CO}^{\rm bkgsub}$ at the cluster centre is higher at 60 pc resolution than 150 pc resolution, which reinforces that YNO clusters coincide with molecular gas peaks. For non-YNO clusters, their radial profiles show a large scatter without significant signal due to the limited sample size. Revisiting this analysis with higher resolution CO maps will be a key next step. 

\subsection{The large scale background CO intensity correlates with cluster properties}
\label{subsec:profile_outer}

\begin{figure*}
\centering
\includegraphics[width=0.42\textwidth]{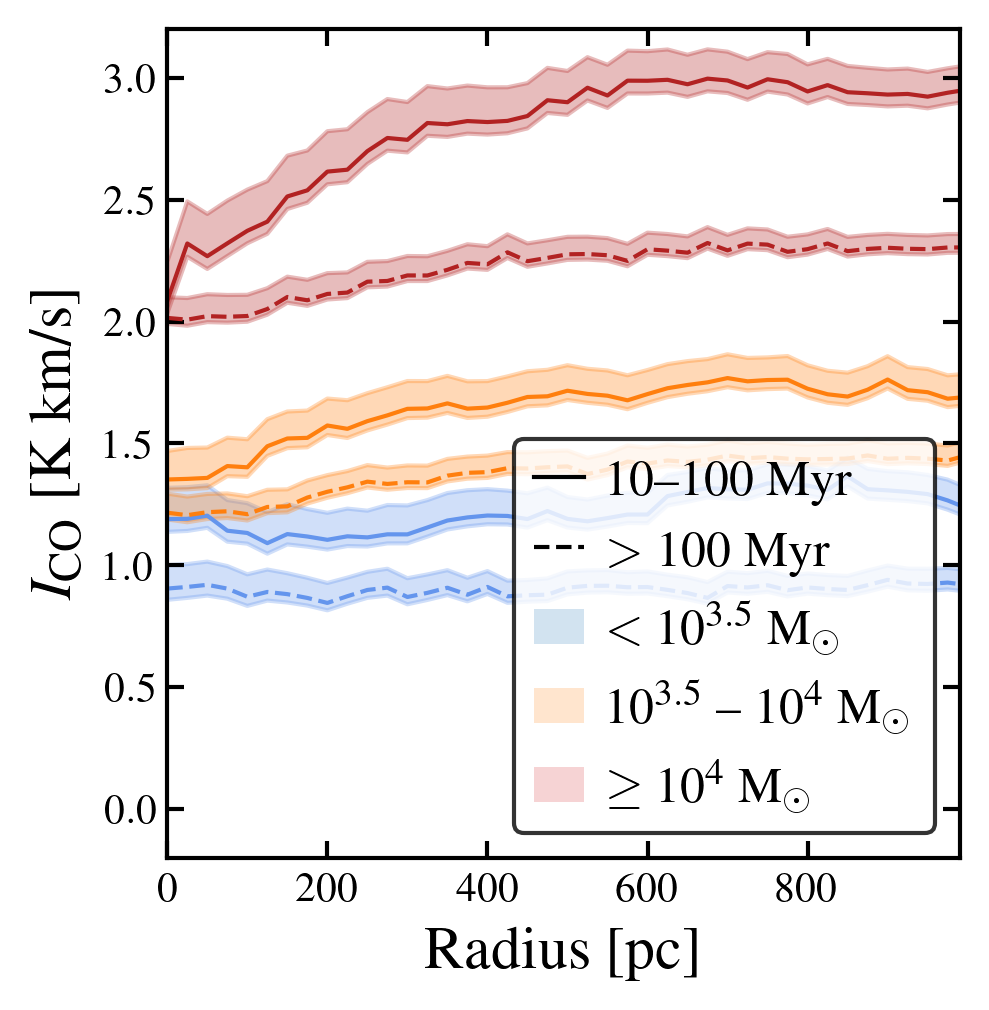}
\includegraphics[width=0.45\textwidth]{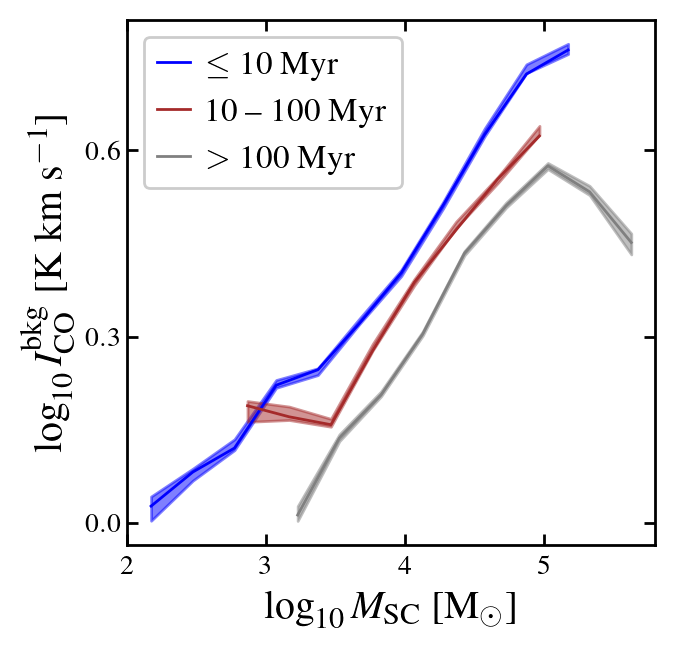}
\caption{(\textit{Left}) Stacked median CO intensity profiles around the old star clusters ($>$ 10 Myr) of different masses. The solid and dashed lines indicate clusters aged 10--100 Myr and $>$100 Myr. Different colours indicate clusters of different mass ranges: $<10^{3.5}$ \solarmass (\textit{blue}), 10$^{3.5}$ -- 10$^4$ \solarmass (\textit{orange}) and $\geq 10^4$ \solarmass (\textit{red}). 
(\textit{Right}) The median of the $I_{\rm CO}$ background within 500 -- 1000 pc annulus for different cluster mass bins (bin width of 0.3 dex) of different age groups: $\leq$10 Myr (\textit{blue}), 10--100 Myr (\textit{brown}) and $>$100 Myr (\textit{gray}). The colour shaded regions indicate the median uncertainty derived from error propagation.   }
\label{fig:bkg_vs_mass}
\end{figure*}

In \S \ref{subsec:Idiff_profile}, we define the local background intensity for individual clusters, $I^{\rm bkg}_{\rm CO}$, as the median CO intensity within the deprojected annulus spanning from 500{-}1000~pc. This large scale $I_{\rm CO}^{\rm bkg}$ correlates with both cluster age and cluster mass, as  shown in Fig. \ref{fig:bkg_vs_mass}. The left panel shows the stacked intensity ($I_{\rm CO}$) profiles for clusters older than 10 Myr of different mass ranges. We see a clear trend that more massive clusters have larger flattened $I_{\rm CO}$ values at 500 -- 1000 pc for a given age range. Meanwhile for a given mass range, clusters of the intermediate age ($10{-}100$~Myr) show a higher $I_{\rm CO}^{\rm bkg}$ than the older $> 100$~Myr clusters. The right panel of Fig. \ref{fig:bkg_vs_mass} directly correlates $I_{\rm CO}^{\rm bkg}$ with cluster mass for the three cluster age groups ($\leq$ 10 Myr, 10--100 Myr and $>$100 Myr). We see a strong positive correlation between $I_{\rm CO}^{\rm bkg}$ and cluster mass, except for old ($>$100 Myr) and massive ($M_{\rm SC} >10^5$ \solarmass) clusters. Younger clusters also have higher $I_{\rm CO}^{\rm bkg}$ than older clusters. 

The $M_{\rm SC}{-}I_{\rm CO}^{\rm bkg}$ correlation presumably stems from the fact that regions of galaxies with high $I_{\rm CO}^{\rm bkg}$ also have high $\Sigma_{\rm SFR}$. For a secularly evolving disk galaxy, the $I_{\rm CO}^{\rm bkg}$ at kpc scales mirrors both the large scale SFR and stellar distribution, so this explanation should hold even out to $> 100$~Myr. Then the implication is that regions with high SFR preferentially form a larger number of massive clusters. Radial migration due to dynamical friction \citep[e.g.][]{lotz_dynamical_2001} is inefficient to create such cluster mass segregation. For a typical cluster in our sample ($\sim10^4$ \solarmass), the dynamical friction timescale can be 500 Gyr, which is much larger than the Hubble time\footnote{A cluster of 10$^7$ \solarmass at a galactic radius of 5 kpc requires $\sim$ 5 Gyr to migrate inwards \citep{forbes_globular_2018}, with the timescale inverse proportional to cluster mass.}.


Preferential formation of more massive clusters in regions of high SFR is in tension with the observation that the cluster mass function appears relatively universal with slope $\approx -2$ \citep[][]{chandar_fraction_2017,krumholz_star_2019, chandar_star_2026}. That would imply an equal distribution of cluster masses per unit star formation regardless of the overall intensity of star formation. There may be stochastic effects at play similar to the sampling of the stellar IMF \citep[e.g.][]{lee_imf_2009}. There may also be other considerations like a link between maximum cluster mass and SFR \citep[e.g.][]{weidner_2005_igimf}, for example because the total amount of dense gas available for cluster formation might set an upper limit of the maximal cluster mass it can form \citep[e.g.][]{mok_mass_2020}. Further exploring any dependence of maximum $M_{\rm SC}$ on environment in PHANGS is important but beyond the scope of this paper. Similarly, observational effects (e.g. the recovery of low mass clusters in dense high SFR regions) will be addressed in future work.


The higher $I_{\rm CO}^{\rm bkg}$ for younger clusters at fixed $M_{\rm SC}$ would naturally result from cluster drift. Given a typical cluster drift velocity of 10 \velu (see \S \ref{subsec:feedback_drift}), it takes a cluster $\sim$100 Myr to travel $\sim 1$ kpc. Since more clusters tend to be born near spiral arms or in the denser inner regions of galaxies, this drift will drift into lower density regions $I_{\rm CO}^{\rm bkg}$ and away from spiral arms. This expectation agrees with the observation that many clusters with age $\leq$10 Myr are located within CO spiral arms (e.g. for NGC 628 see Fig. \ref{fig:ngc628_maps}), while intermediate and older clusters have a more azimuthally smooth distribution. It also agrees with clustering analyses by \citet[][]{bastian_spatial_2009, grasha_spatial_2019} in which older clusters have distributions similar to the stellar disk. Stellar feedback,  which will include SNe by the later time bins discussed here, will amplify this effect, disperse and reshape dense gas concentrations. Thus over time we expect to see a continuous drop in $I_{\rm CO}^{\rm bkg}$ as clusters get older and lose their link to their birth environment.

\subsection{The impact of cluster drift on the observed cluster-gas dissociation}
\label{subsec:feedback_drift}

\begin{figure}
    \centering
    \includegraphics[width=0.45\textwidth]{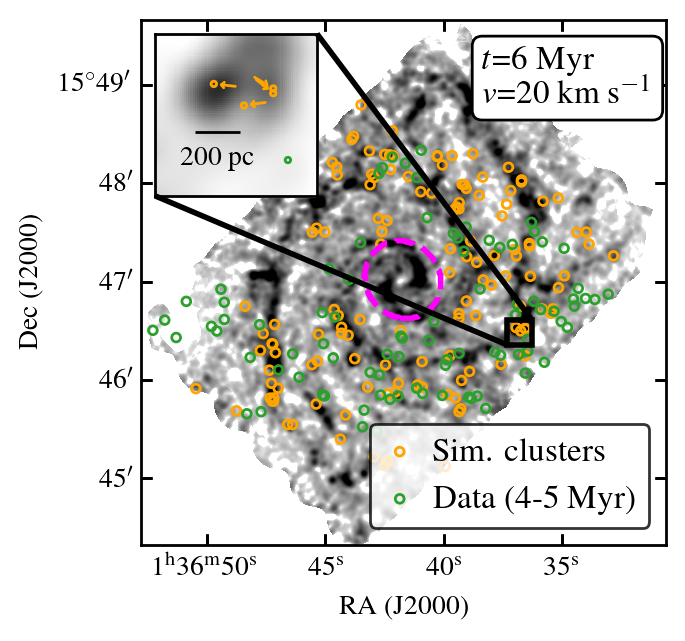}
    \caption{CO~(2-1) map of NGC 628 overlaid with one set of simulated cluster locations at a snapshot of 6 Myr (\textit{orange circles}) calculated assuming a drift velocity of 20 \velu. The green circles indicate real, observed non-YNO clusters with ages of 4--5 Myr. The orange arrows in the zoom-in view show the route traveled by the simulated clusters.  }
    \label{fig:ngc0628_sim_drift_map}
\end{figure}

\begin{figure*}
\centering
\includegraphics[width=0.9\textwidth]{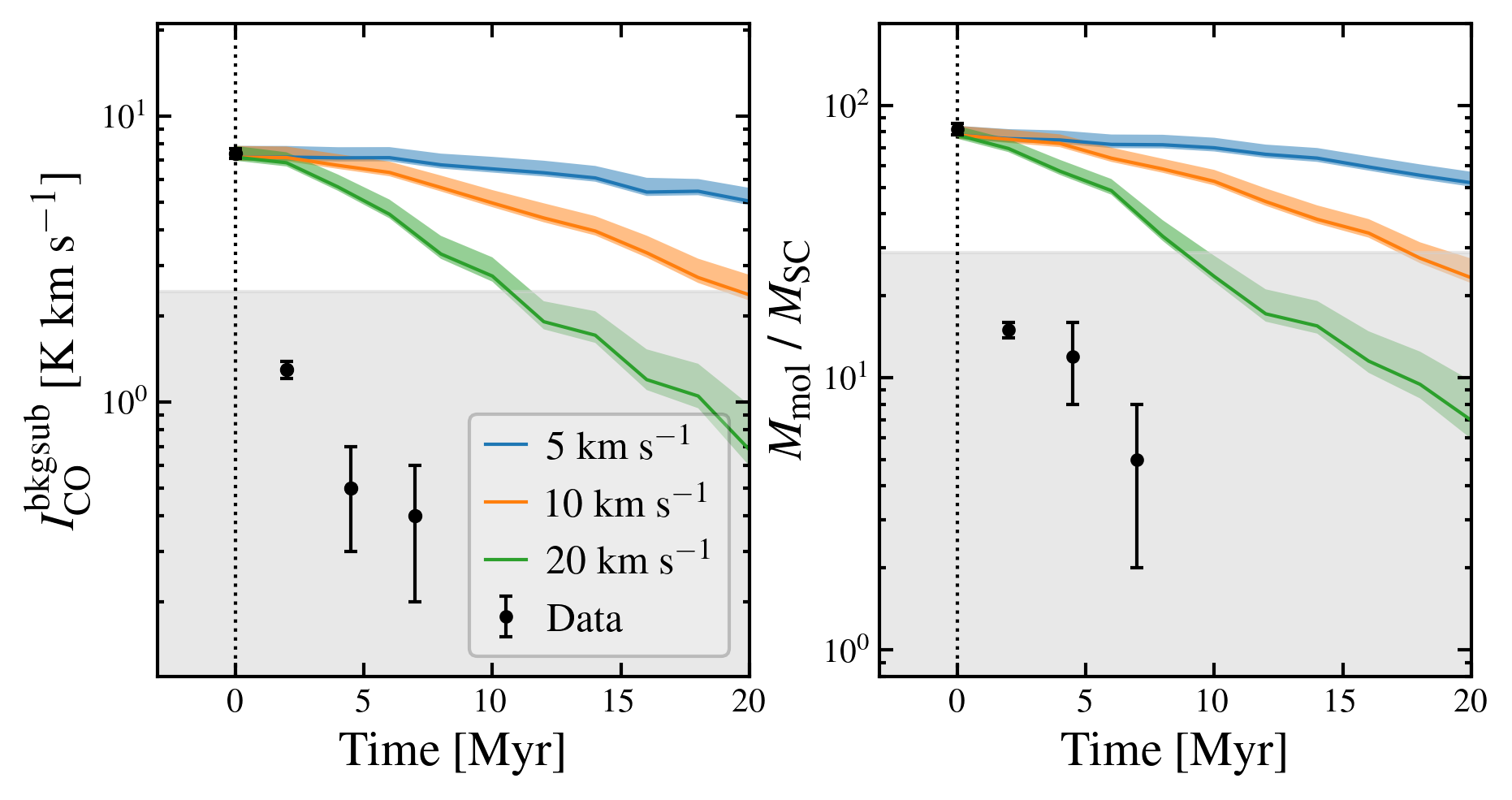}
\vspace{-1\baselineskip}
\caption{(\textit{Left}) $I_{\rm CO}^{\rm bkgsub}$ and (\textit{right}) $M_{\rm mol}/M_{\rm SC}$ for simulated clusters drifting from the initial position of YNO clusters. Coloured lines and shaded regions indicate the median and median uncertainty of simulations with different cluster drift velocities: 5 \velu (\textit{blue}), 10 \velu (\textit{orange}) and 20 \velu (\textit{green}). Black points and error bars indicate measured values from the data. The data point at 0 Myr indicates YNO clusters, which we use as initial condition for our cluster drift simulation. 
The light-gray shaded regions indicate the typical 3$\sigma$ area for individual measurements. The exercise demonstrates that cluster drift alone cannot result in the observed decrease in $I_{\rm CO}^{\rm bkgsub}$ as a function of cluster age.}
\label{fig:cluster_drift}
\end{figure*}


One potential explanation for the observed cluster--gas dissociation is that star clusters drift away from their natal molecular clouds \citep{koda_on_2023}. We model this drift process to estimate the characteristic timescale over which it becomes effective. We consider YNO clusters as a population fully coupled with molecular clouds and set them as the starting point of our cluster drift model. We then give each YNO cluster ($M_{\rm SC}\geq10^{3.5}$ \solarmass) three fixed velocities (5, 10 and 20 \velu) towards a random location within the deprojected galactic plane, and let it evolve for 20 Myr. For every 2 Myr, we record cluster positions and measure their corresponding $I_{\rm CO}^{\rm bkgsub}$ and $M_{\rm mol}/M_{\rm SC}$ at that snapshot. Fig. \ref{fig:ngc0628_sim_drift_map} shows an example snapshot of NGC 628 at 6 Myr with drift velocity of 20 \velu. 

Our model depends sensitively on the assumed cluster drift velocity. Galactic studies with Gaia \citep[e.g.][]{soubiran_open_2018} show a typical velocity dispersion of $\sim10$~\unit{\kilo\metre\per\second} for young open clusters (age $<60$~Myr), while the older clusters exhibit a larger velocity dispersion of $\sim20$~\velu. Our Solar System also has a velocity of 20~\velu relative to the local ISM \citep{frisch_interstellar_2011}. For nearby galaxies in the Local Group, such as M33, spectroscopic measurements \citep[e.g., ][]{chandar_kinematics_2002} show a similar velocity dispersion of 17~\velu for clusters younger than 100~Myr. \citet{peltonen_clusters_2023} performed GMC-cluster cross correlation in M31 and M33 and also estimate a similar drift velocity of 5-10~\velu. In summary, our assumed cluster drift velocity of 5 -- 20~\velu is slightly higher than the typical drift velocity measured for young star clusters, and so may overestimate the impact of cluster drift on the spatial cross-correlation.

The origin of the cluster drift relative to their parent cloud is still under exploration. A widely accepted scenario is that the drift velocity comes from the random motion within the parent GMC. In this scenario, we expect cluster drift velocity to be similar to the GMC velocity dispersion. GMCs in PHANGS galaxies have a median velocity dispersion of 5~\velu \citep[e.g., ][]{rosolowsky_giant_2021, sun_star_2023}, which translates to a 2D velocity dispersion of 7~\velu assuming isotropy ($\sigma^2_{v, \mathrm{2D}} = 2 \sigma^2_{v, \mathrm{1D}}$), which is also comparable to cluster drift velocity estimates from the literature.

In Fig. \ref{fig:cluster_drift}, we show the simulated $I_{\rm CO}^{\rm bkgsub}$ and $M_{\rm mol}/M_{\rm SC}$ as a function of drifting timescale. With our maximum velocity of 20 \velu, it takes $\sim$10 Myr for the median values of these quantities to drop below our individual 3$\sigma$ CO detection limit. It would take $\sim$20 Myr for cluster drift to drive $I_{\rm CO}^{\rm bkgsub}$ to the values observed for non-YNO clusters with modelled ages of 1--3 Myr. Because the typical cluster drift velocity is expected to be lower than 20 \velu , we would expect even longer timescales to be needed for cluster drift to drive the observed cluster-molecular gas disassociation. This supports the idea that the much shorter dissociation timescale in our observations (4--6 Myr, \S \ref{subsec:Ico_hist_age}) is primarily due to stellar feedback and cloud destruction, instead of cluster drift.

\section{Comparison with mock clusters}
\label{sec:mock_control}

To investigate the origin of the central depressions in $I_{\rm CO}^{\rm bkgsub}$ observed towards intermediate age and old clusters (Fig. \ref{fig:stacked_profile} and \ref{fig:bkg_vs_mass}) and the origin of the large scatter in the $I_{\rm CO}^{\rm bkgsub}$ vs. $M_{\rm SC}$ relation (Fig. \ref{fig:Idiff_mass_young}), we construct mock cluster catalogues. We repeat our measurements on these catalogues and compare to the observed results.

\subsection{Construction of mock cluster control}
\label{subsec:mock_cats}

\begin{figure*}
\centering
\includegraphics[width=\textwidth]{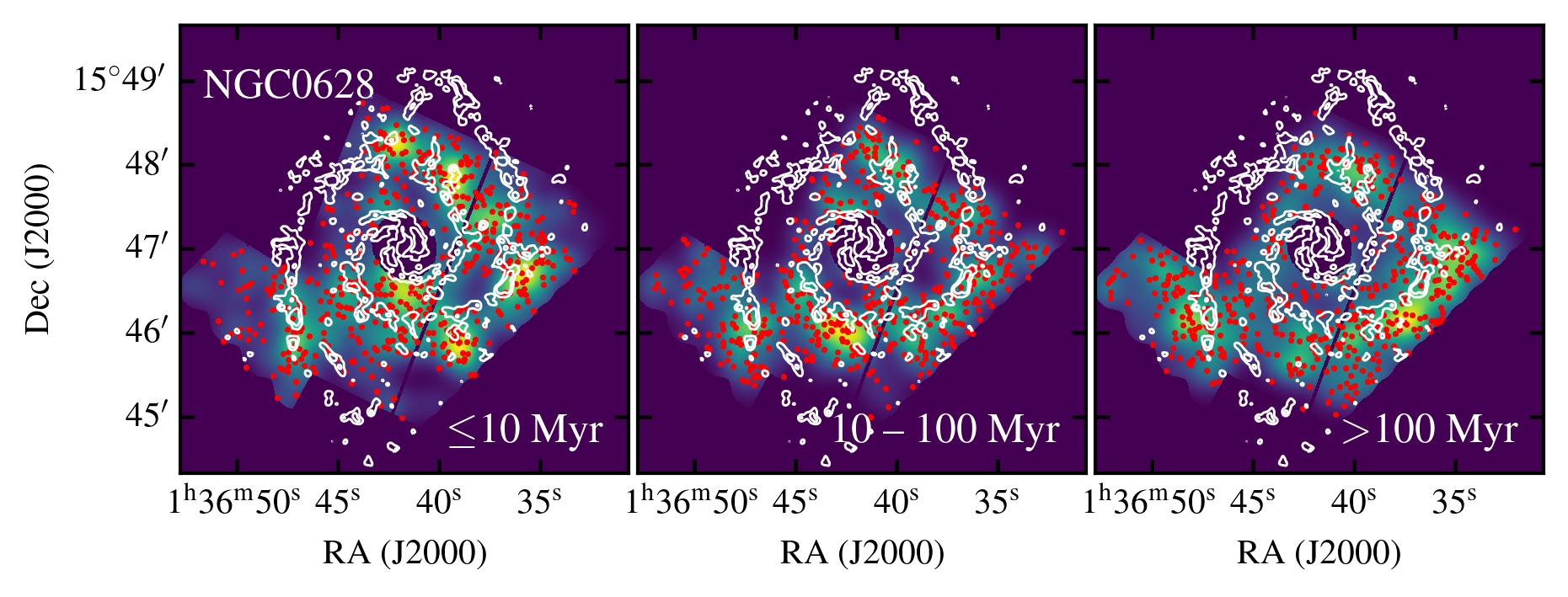}
\caption{Mock clusters (\textit{red}) overlaid on the probability density fields (colour map) constructed from the kpc-scale cluster number density distribution for three cluster age groups: (\textit{left}) clusters younger than 10 Myr, (\textit{middle}) clusters 10 -- 100 Myr and (\textit{right}) clusters older than 100 Myr. The white contours show CO(2-1) integrated intensity with intensity of 4 and 10 K \velu.   }
\label{fig:ngc0628_density_fields}
\end{figure*}

We require a two-dimensional probability density field to generate the mock cluster positions. At $\gtrsim$~1~kpc scales the spatial distribution of star-forming regions is largely set by the galaxy’s radial profile and morphological features such as spiral arms and bars \citep{menon_dependence_2021, he_structure_2026}. To capture this large-scale structure, we construct our probability density field by smoothing the number of clusters per unit area in the real catalogue to a resolution of 1~kpc.

Mechanically, we create a map that covers the overlapping footprint of HST and ALMA observation coverage and excludes the centre regions, which we do not analyse here. For each pixel that contains a star cluster within this footprint, we add a flux value of 1. We then smooth this cluster location map by a Gaussian kernel with a major full width half maximum (FWHM) of 1~kpc (corrected for the inclination and position angle of galaxies). The result is a two dimensional probability density field that reflects the large-scale distribution of the clusters. We repeat this exercise to create probability maps for clusters in different age bins.

We then draw a sample of mock sources from this field. We set the number of mock clusters to be the same as the real source numbers ($N_{R}=N_{D}$)\footnote{We also test $N_{R}=100 N_{D}$ and see no significant difference in measured radial profiles.}. For each galaxy, we generate three mock catalogues, one drawn from the large-scale distribution of clusters with age $\leq$10 Myr\footnote{Sub-groups of clusters with age $\leq$10 Myr may have different large-scale spatial structure. However, the limited sample size of clusters in individual galaxies makes it challenging to build robust large-scale density fields for these sub-groups.}, one for 10 -- 100 Myr clusters, and one for $>$100 Myr clusters. Fig. \ref{fig:ngc0628_density_fields} shows an example of this process for NGC 628 ).  

\subsection{Stacked $I_{\rm CO}^{\rm bkgsub}$ profiles of mock clusters}
\label{subsec:profile_dip}

\begin{figure}
    \centering
    \includegraphics[width=0.5\textwidth]{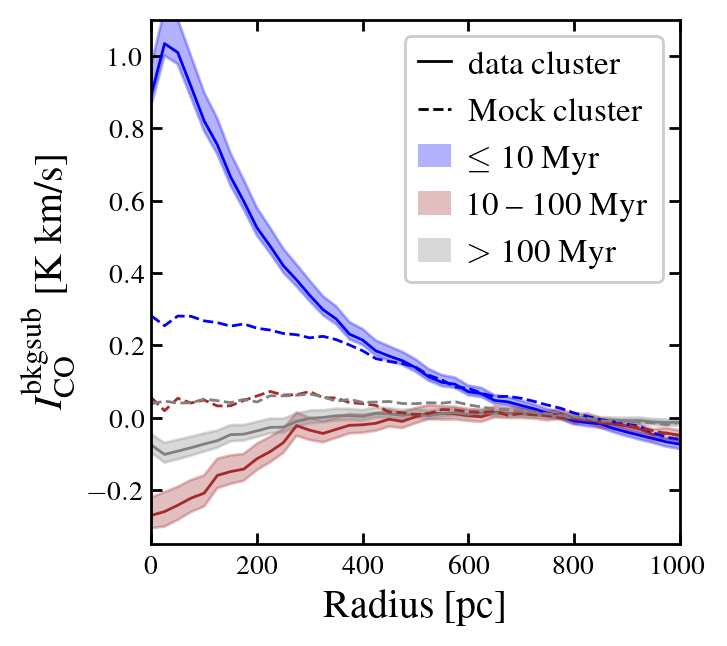}
    \caption{The radial profile of $I^{\rm bkgsub}_{\rm CO}$ for observed clusters and their associated mock clusters. Solid lines indicate the stacked profile for clusters of different age groups: $\leq$10 Myr (\textit{blue}), 10 -- 100 Myr (\textit{brown}) and $>$ 100 Myr (\textit{gray}). Dashed lines indicate the stacked radial profiles of mock clusters for the same age groups. Mock clusters for cluster with age $> 10$~Myr show flat radial profiles while the real clusters show a significant central dip. This might be due to neighbouring clumpy CO structures or bias in the cluster catalogues.}
    \label{fig:profl_dip}
\end{figure}

We measure the stacked radial profiles of $I_{\rm CO}^{\rm bkgsub}$ for mock clusters and compare them with those of observed clusters, as shown in Fig. \ref{fig:profl_dip}. The $I_{\rm CO}^{\rm bkgsub}$ profiles for mock clusters align well with those of observed clusters at scales above 500 pc. For clusters older than 10 Myr, the mock profile is flat with values very close to zero. 

For clusters younger than 10 Myr, the mock profile has a peak value of $\sim$0.2 K~km~s$^{-1}$, much lower than the observed value. This is expected, because the kpc-resolution density maps we draw the mocks from will preferentially place young clusters around spiral arms but lack information on the small scale CO distribution. This highlights that the CO~(2-1) excess observed near the YNOs is indeed a local effect.

For clusters older than 10 Myr, we find a weak but significant central dip in the observed $I_{\rm CO}^{\rm bkgsub}$ profile (Fig. \ref{fig:profl_dip}), while the mock profiles remain flat. This implies that the observed clusters are preferentially distributed in locally CO-depleted regions. That is, the dip in the profile does not reflect the large-scale structure of the galaxy, which would be reflected in our mock profiles. 

One explanation for the central dip is that intermediate age clusters tend to be near CO emission but not coincident with it. For example, clusters will drift away from spiral arms, which will move through the disk as a density wave \citep[e.g.][]{williams_phangs_2022}. Or clusters will simply be born with some velocity dispersion with respect to their parent cloud. Alternatively clusters may disperse their parent clouds but remain near neighboring clouds due to the intrinsic clustering of molecular gas \citep[e.g.][]{he_structure_2026}. This scenario could explain the stronger dip for clusters with intermediate age of 10--100 Myr compared to clusters older than 100 Myr, 
as well as the higher background levels for those clusters (Fig. \ref{fig:bkg_vs_mass}). 
We expect clusters with intermediate age of 10 -- 100 Myr are still in the process of drifting away from the gas-rich regions \citep[typical cluster drift timescale is $\sim$15 Myr, see \S \ref{subsec:feedback_drift} and][]{peltonen_clusters_2023}, while clusters older than 100 Myr are more widely distributed across the whole galactic disk. 

Alternatively, the central dip might be caused by the incomplete detection of old clusters in CO-rich regions, where dust extinction is high. We would expect this extinction-related bias to be age-independent while the dynamical or evolutionary explanations will depend on age. Therefore we measured the $I_{\rm CO}^{\rm bkgsub}$ profile for a subset of clusters older than 1 Gyr. In this case, we find that the difference in central $I_{\rm CO}^{\rm bkgsub}$ between data and mock changes to 0.08 K~km~s$^{-1}$, slightly higher $I_{\rm CO}^{\rm bkgsub}$ for the data. This is also consistent with the stronger central dip observed in $10{-}100$ Myr clusters compared to $> 100$~Myr clusters. Based on this, we favour an evolutionary explanation. We note that a more thorough and robust test would require an environment-dependent completeness correction, such as the C-4 framework of \citep{tang_environmental_2026} to unambiguously separate the two factors. In the near future, we also expect JWST-based cluster catalogues that are much more robust to extinction to address the issue of completeness more thoroughly. 

\subsection{Origin of the scatter in the CO vs. cluster mass relations}

\begin{figure*}
\centering
\includegraphics[width=0.9\textwidth]{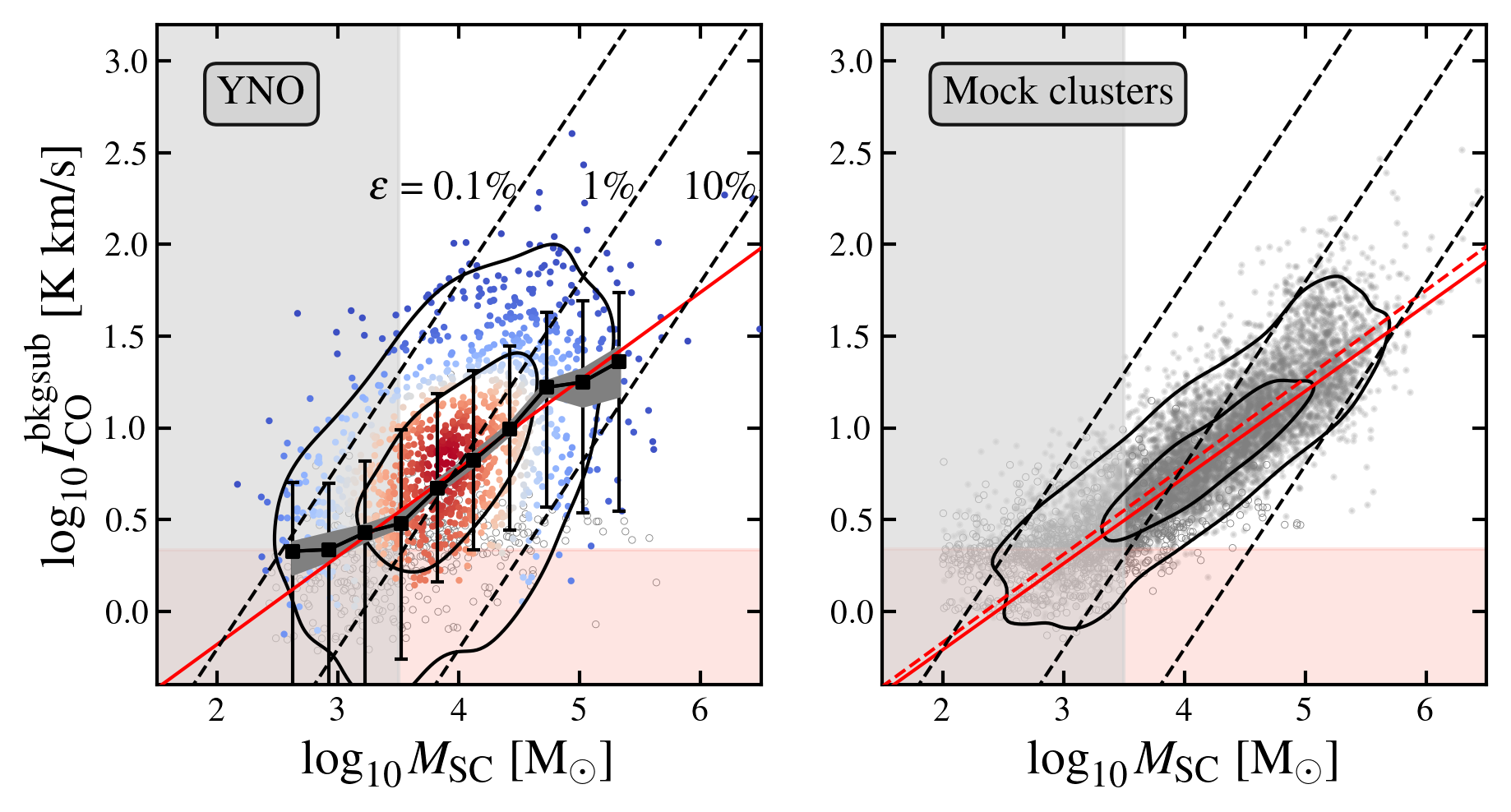}
\caption{(\textit{Left}) The background subtracted $I_{\rm CO}$ versus cluster mass for YNO clusters (same notation as Fig. \ref{fig:Idiff_mass_young}).  The contours enclose the 39.95\% and 86.47\% of positive points.  
(\textit{Right}) The background subtracted $I_{\rm CO}$ versus cluster mass for artificially injected mock clusters following the fitted relation between cluster mass and $I^{\rm bkgsub}_{\rm CO}$ (Eq. \ref{eq:Idiff_vs_mass}, see text for details). The red dashed line is the \texttt{linmix} power-law fit to the mock data. 
We see that the large scatter we observe in $I^{\rm bkgsub}_{\rm CO}$ vs $M_{\rm SC}$ relation is partially due to the clumpy CO structure that induces scatter in background subtracted values.  }
\label{fig:obs_vs_mock_scatter}
\end{figure*}

The correlations in Fig. \ref{fig:Idiff_mass_young} and histograms in Fig. \ref{fig:cluster_co_intensity} show significant scatter, and although the correlations are formally significant most trends emerge only after averaging a large number of clusters. Our background subtraction method will be affected by the noise in the CO~(2-1) maps and the spatially uneven background in the clumpy CO maps. 


To test the scatter expected due to noise and CO substructure, we inject artificial Gaussian sources into the CO~(2-1) maps at the mock catalogue positions for young clusters (age $\leq$ 10 Myr). The peak value of injected $I_{\rm CO, 150pc}$ is set based on the kpc-resolution CO intensity value at the cluster position following the relation \citep[adapted from][]{sun_molecular_2022}
\begin{equation}
\frac{I_{\rm CO, 150 pc}}{\rm K\ km\ s^{-1}} =  2.5 \left(\frac{I_{\rm CO, 1 kpc}}{\rm K\ km\ s^{-1}}\right)^{0.77}
\end{equation}
This relation is solely used to set the artificial $I_{\rm CO, 150pc}$ to a reasonable range. It does not contribute to the scatter in the final relation. After drawing $I_{\rm CO, 150 pc}$ we set the corresponding expected cluster mass assuming that cluster formation follows perfectly Eq. \ref{eq:Idiff_vs_mass} (we remove mock clusters where the scaling relation would give a $M_\mathrm{SC}<100$\,\solarmass ). Then we randomly add noise to the simulated $I_{\rm CO, 150pc}$. Finally we perform the same background subtraction used on the data to obtain the mock observed $I_{\rm CO}^{\rm bkgsub}$. 


We show the comparison between observed $I_{\rm CO}^{\rm bkgsub}$ vs $M_{\rm SC}$ relation for YNO clusters and the mock relation in Fig. \ref{fig:obs_vs_mock_scatter}. The observed relation has a 1$\sigma$ scatter of 0.49 dex around our fitted power-law relation according to the \texttt{linmix} fit results. For mock clusters, a \texttt{linmix} fit yields a similar power-law relation but with scatter of 0.21 dex. This suggests $\sim$40\% of the observed scatter can be attributed to these systematic effects. 

The remaining scatter in the observed correlation could have several physical origins. First, our CO(2-1) maps at 150~pc resolution only trace the mass of molecular gas over a large region, while star clusters form out of dense clumps \citep[e.g.][]{mckee_theory_2007, heyer_molecular_2015}. Therefore, we might expect a stronger correlation with higher resolution maps that can isolate the locally star-forming gas. 
Second, the cluster mass formed out of a given amount of molecular gas mass may have intrinsic scatter either due to varying local conditions that affect star formation or simply stochastic processes during the core and star formation process.
Relevant to this, analysing Milky Way observations \citep[e.g.,][]{lee_observational_2016} also found a large scatter in the inferred star formation efficiency that cannot be explained from turbulence theory \citep[e.g., ][]{krumholz_general_2005}. 

\section{Discussion \label{sec:discussion}}

\subsection{Feedback timescales}
\label{subsec:feedback_timescale}


Our measurements constrain the ``cluster feedback timescale,'' which we define as the period when clusters and their natal gas coexist. They show that stellar feedback has already been effective at 1--3 Myr and that most further gas clearing completes during the 4--6 Myr window. This is consistent with both previous cluster-focused studies and statistical analysis comparing tracers of gas and star formation, especially H$\alpha$. 

Previous star cluster studies using HST data alone \citep[e.g., ][]{hollyhead_studying_2015, grasha_connecting_2018, hannon_h_2019, hannon22} show that star clusters coming out of an embedded phase become optically visible after $\sim4$~Myr. Studies 
\citep[e.g., ][]{whitmore_alma_2014, he_embedded_2022, sun_hidden_2024} that combine the HST data and radio continuum data, which probe star clusters during their embedded phase, also give similar emerging timescales of 3 -- 6~Myr. Recent studies that inspect ISM morphology around star clusters \citep[e.g.][]{whitmore_improving_2023} and the spatial correlation between star clusters and GMCs  \citep[e.g.][]{grasha_spatial_2019, turner_phangs_2022, peltonen_clusters_2023} for individual spiral galaxies also give similar feedback timescales of $\sim$ 4~Myr.

Besides star cluster studies, the relative spatial distribution of H$\alpha$ and CO \citep[the so-called ``tuning-fork analysis''][]{schruba_tuningfork_2010} has also been extensively used to infer the feedback timescale and cloud lifetime \citep[e.g., ][]{chevance_lifecycle_2020,kim_on_2021, kim_environmental_2022, kim_phangs_2023, ramambason_duration_2026}. In their definition, feedback timescale refer to the co-existence timescale between CO and H$\alpha$, which has a similar but not identical physical meaning to our definition. These studies find a similar feedback timescale of 1 -- 5~Myr for a large sample of PHANGS galaxies at GMC scales (60--150 pc). Directly comparing the areal ratio between CO-H$\alpha$ overlap and H$\alpha$-only regions yields similar feedback timescale for PHANGS galaxies \citep{schinnerer_gas_2019, pan_gas-star_2022}. 

Compared to previous star cluster studies, our work provides a more direct measurement and results with more statistical power and better time sensitivity thanks to the unprecedented size of the PHANGS-HST cluster sample. Compared to the H$\alpha$+CO methods, our use of SED-constrained ages should represent an improvement over simply adopting a fiducial timescale for H$\alpha$, though our YNO population does use H$\alpha$ detection as part of the classification. The main point that we emphasize here, however, is good agreement between our work and the recent literature.

\subsection{Cluster mass to gas mass ratio}
\label{subsec:formation_efficiency}

Our observed cluster mass to gas mass ratio, $\epsilon_{\rm obs}$ for the young YNO clusters constrains the efficiency of clusters formed out of molecular clouds. Studies that directly compare the cluster mass function and the GMC mass function \citep[e.g., ][]{mok_mass_2020} suggest a similar ratio of $\sim$1\% as our measured ratio for YNO clusters. 

From a theoretical perspective, assuming that $\epsilon_{\rm obs}$ of YNO clusters is regulated only by the formation process ($\epsilon_{\rm feedback}=0$, Eq. \ref{eq:efficiency}), our measured efficiency $\epsilon_{\rm obs}$ consists of two terms
\begin{equation}
\epsilon_{\rm obs} \approx \epsilon_{\rm form} = \epsilon_{\rm SFE} \times \Gamma~,
\end{equation}
where $\epsilon_{\rm SFE}$ is the star formation efficiency that measures what fraction of gas converts to stars and $\Gamma$ is the bound fraction that measures what fraction of stars are formed in star clusters. 

Observations of GMCs in the Milky Way and nearby galaxies \citep{murray_star_2011, lee_observational_2016, ochsendorf_what_2017, zhou_most_2025} show that $\epsilon_{\rm SFE}$ ranges from 0.1\% -- 10\%, with a fiducial value of 1\% that is often adopted in theoretical turbulence models \citep[e.g.][]{krumholz_general_2005, hennebelle_analytical_2011}. Simulations of individual GMCs \citep[e.g., ][]{kim_giant_molecular_2021} also reproduce these $\epsilon_{\rm SFE}$ values within a typical star formation timescale ($\sim$ 3 Myr). Meanwhile, the cluster formation efficiency $\Gamma$ is more uncertain. \citet{kruijssen_fraction_2012} propose a theoretical model suggesting $\Gamma$ is dependent on SFR or SFR surface density, $\Sigma_{\rm SFR}$. For typical PHANGS galaxies with $\Sigma_{\rm SFR}$ of 10$^{-3}$ -- 10$^{-1}$ M$_{\odot}$~yr$^{-1}$~kpc$^{-2}$ the predicted $\Gamma$ for this model ranges from 1\% to 30\% \citep[][Chandar R. in prep.]{chandar_tale_2023}. However, recent simulations \citep[e.g.][]{li_star_2018, dinnbier_do_2022, grudic_dynamics_2022} suggests a much shallower dependence of $\Gamma$ on $\Sigma_{\rm SFR}$, with a predicted $\Gamma$ of 10\% to 40\% depending on the model choice for a typical PHANGS $\Sigma_{\rm SFR}$ of 0.01 M$_{\odot}$~yr$^{-1}$~kpc$^{-2}$.

Recent observations \citep[][]{chandar_tale_2023, chandar_star_2026} suggests a constant $\Gamma$ relative to $\Sigma_{\rm SFR}$ with a median value of $\sim$20\%. Using the fiducial $\epsilon_{\rm SFE}$ of 1\% and the fiducial $\Gamma$ of 20\%, we expect a typical $\epsilon_{\rm obs}$ of 0.2\%, which is below the 1.3\% value we find for YNO clusters. One possible explanation for the discrepancy is that literature measurement of $\Gamma$ are generally for clusters younger than 10 Myr. Our YNO clusters can be much younger than this and we might expect a higher $\Gamma$ for YNO clusters if a significant fraction of clusters are disrupted during the first 10~Myr \citep[as happens later, e.g., see][]{krumholz_star_2019, chandar_star_2026}. Alternatively, our assumption of $\epsilon_{\rm feedback}=0$ for YNO clusters may not hold. These clusters are already exposed and optically visible, suggesting a certain degree of gas clearing. In this case, we may overestimate the actual $\epsilon_{\rm form}$ by neglecting the effect of stellar feedback. A natural next step is to perform the same analyses for more embedded clusters, such as 3.3\micron \citep{rodriguez_star_2025}, 10\micron and 21\micron \citep{hassani_hidden_2026} sources to probe the early evolution of $\epsilon_{\rm obs}$. 

Note that our observed $\epsilon_{\rm obs}$ does not take into account multiplicity in cluster formation and only considers clusters with $M_{\rm SC} \geq 10^{3.5}$~M$_\odot$. When several clusters, including clusters below our mass threshold, form within the same beam-sized region, we will underestimate the $\epsilon_{\rm form}$ of that region. Accounting for multiple clusters forming from a single gas concentration is likely to raise $\epsilon_{\rm form}$. Since this has the opposite sense of the bias imposed by early feedback, we do not label $\epsilon_{\rm form}$ as either an upper or lower limit, and just note these effects as important directions for future work.

We also note that $\epsilon_{\rm SFE}$ measurements vary in the literature. For comparison, also using the PHANGS data sets, \citet{pathak_linking_2025} recently estimated $\epsilon_{\rm form}$ by comparing statistical estimates of the $M_{\rm mol}$ ``missing'' from the sites of \textsc{Hii} regions to the $M_{\rm SC}$ of stars needed to power the \textsc{Hii} region. They found typical  $\epsilon_{\rm form} \sim 7{-}9\%$, which is higher than the $\epsilon_{\rm form} \sim 1\%$ for YNOs here. The difference likely reflect the different methods used and highlights the uncertainty in both results. \citet{pathak_linking_2025} focus on statistical estimates of dispersed $M_{\rm mol}$, while we measure the full local $I_{\rm CO}^{\rm bkgsub}$. Both approaches attempt to isolate the gas mass associated with star formation, but the \citet{pathak_linking_2025} approach is more aggressive and the one here is more conservative. A natural next step will be to apply our method also to the \citet{groves_phangs_2023} PHANGS \textsc{Hii} regions to compare results. Along this direction, several other star-forming tracers, such as supernovae remnants \citep{li_discovery_2024}, red supergiants \citep{sarbadhicary_2026} and WR stars (Liang F. in prep.), can give us a full census of star formation products, and help us constrain the energy and momentum input from different stellar populations. 

Finally, we find that our measured $\epsilon_{\rm obs}$ for YNO clusters has a positive correlation with $M_{\rm SC}$ and $I_{\rm CO}^{\rm bkgsub}$ (Fig. \ref{fig:Idiff_mass_young} and \ref{fig:obs_vs_mock_scatter}). Assuming that $\Gamma$ is constant, this suggests a higher $\epsilon_{\rm SFE}$ for more massive GMCs. This is consistent with some predictions from numerical simulation \citep[e.g.][]{hopkins_self_2011, grudic_when_2018}. However, our results are in a slight tension with Milky Way observations \citep[e.g.][]{murray_star_2011, lee_observational_2016}, which suggests a lower $\epsilon_{\rm SFE}$ for more massive GMCs \citep[though the situation is more ambiguous in recent work by][where the gravitational free-fall time remains approximately constant for most massive clouds]{evans_star_2022,evans_star_2026}. Most of these studies begin their analysis from identified GMCs, rather than beginning with clusters as we do here. An important next direction for extragalactic work will be to measure the present mass of YNO clusters at all molecular gas peaks, but this requires better resolution than PHANGS--ALMA to yield results analogous to Milky Way studies.

\section{Conclusions}
\label{sec:conclusion}

We use the PHANGS--ALMA CO~(2-1) maps to measure CO intensity and estimate molecular gas mass at the sites of the star clusters identified from PHANGS-HST by \citet{maschmann_phangs-hst_2024,thilker_phangs-hst_2025}. The cluster catalogues are sensitive to star clusters with stellar mass $M_{\rm SC} > 10^{3.5}$~M$_\odot$ out to ages of $\gtrsim 20$~Myr (Fig. \ref{fig:mass_vs_age}) and each cluster has an associated age based on SED fitting. Our main conclusions are:

\begin{enumerate}
    \item After subtracting a local background, we find that for clusters with $M_{\rm SC} \geq 10^{3.5}$~M$_\odot$, $I_{\rm CO}^{\rm bkgsub}$ at the cluster centre follows an apparent evolutionary sequence (Fig. \ref{fig:cluster_co_intensity}). Young nebular object (YNO) clusters have the highest median $I_{\rm CO}^{\rm bkgsub}$ of 7.4 K~km~s$^{-1}$. For non-YNO clusters, the youngest population, with age 1--3 Myr, has the highest median $I_{\rm CO}^{\rm bkgsub} = 1.3$ K~km~s$^{-1}$. The median $I_{\rm CO}^{\rm bkgsub}$ declines for bins of clusters with older ages, reaching zero for clusters older than 9--10 Myr. This suggests a short $\sim$4 Myr timescale over which star clusters become disassociated from their natal molecular clouds. 
    
    \item Based on the measured $I_{\rm CO}^{\rm bkgsub}$, we estimate the total molecular gas mass $M_{\rm mol}$ within a 150 pc beam around individual clusters and calculate the gas-to-cluster mass ratio $M_{\rm mol}/M_{\rm SC}$. $M_{\rm mol}/M_{\rm SC}$ also decreases as clusters get older. YNO clusters also have the highest $M_{\rm mol}/M_{\rm SC}$ of 82, which suggests $\sim$1.2\% of gas mass converts to cluster mass during cluster formation. $M_{\rm mol}/M_{\rm SC}$ decreases to 15 for non-YNO clusters of 1--3 Myr, which suggests a significant amount of gas clearing during this short period that is probably due to early stellar feedback.
    
    \item Considering individual clusters, $I_{\rm CO}^{\rm bkgsub}$ at the cluster centre has a significant positive correlation with cluster mass for YNO clusters (Spearman of $r$ of 0.38, p-value of 4e-39). The correlation becomes weaker for non-YNO clusters of 1--3 Myr (Spearman $r=0.16$, p-value of 4e-16), and becomes null for clusters older than 6 Myr. For YNO clusters, the slope of the $M_{\rm mol}$ vs. $M_{\rm SC}$ relationship is below linear. This implies a higher formation efficiency for high-mass clusters.   

    \item To simulate the impact of cluster drift on the observed cluster-gas dissociation, we model cluster drift using the YNO clusters as initial conditions (Section \ref{subsec:feedback_drift}). With cluster drift alone at a maximal velocity of 20 \velu, it takes $\sim$20 Myr for $I_{\rm CO}^{\rm bkgsub}$ and $M_{\rm mol}/M_{\rm SC}$ to reduce to the values that we observe for non-YNO clusters with ages 1--3 Myr. This strongly suggests that the early cluster-gas dissociation that we observe is mainly due to stellar feedback rather than cluster drift.  
    
    \item Our stacked CO radial intensity profiles generally flatten to a stable background level at 500 -- 1000 pc. This background level has a positive correlation with cluster stellar mass for both young (age$\leq$ 10 Myr), intermediate (age at 10--100 Myr) and old (age greater than 100 Myr) clusters. The reason for this correlation remains uncertain, but the observation that more massive clusters appear preferentially associated with regions of high molecular gas surface density is in some tension with a universal cluster mass function.

    \item We stack the background-subtracted CO intensity ($I_{\rm CO}^{\rm bkgsub}$) profiles for clusters less and more massive than 10$^{4}$ \solarmass. We find that massive clusters ($M_{\rm SC}>10^4$ \solarmass) of 4 --5 Myr still show central gas concentrations, though with large scatter, while older clusters show central dips in their stacked profiles. To understand the origin of this signal, we generate mock clusters with locations that mirror the large-scale ($>$kpc) cluster spatial distribution but with no knowledge of the small-scale CO distribution. The mock clusters yield flat $I_{\rm CO}^{\rm bkgsub}$ profiles with no central dip. This suggests that the older clusters have a slight spatial preference to appear CO-poor regions. For younger clusters this might result from supernova feedback. For older clusters this reflect the clumpiness of the CO distribution (e.g., bar streams and spiral arms) or some selection effect related to extinction. 
    
    \item We also use our mock clusters to investigate the origin of the scatter in the $I_{\rm CO}^{\rm bkgsub}$ vs $M_{\rm SC}$ relation for YNO clusters. We quantify the impact of noise in the CO (2-1) data and spatial variation of the CO (2-1) background and find these uncertainties accounts for 40\% of the observed scatter. The remaining scatter presumably reflects the limitations of tracing the molecular gas at 150 pc resolution and physical scatter in the cluster formation efficiency, early feedback efficiency, and sampling of the IMF. 
\end{enumerate}

Our method can be easily applied to simulations that model cluster formation. Since such simulations can assess cluster formation efficiency and can track drift directly, this would be a natural way to test our interpretation that early feedback drives the measured age dependence of $I_{\rm CO}^{\rm bkgsub}$ and $M_{\rm mol}/M_{\rm SC}$. Another logical next step will be to apply these methods to higher-resolution CO maps. This will better spatially resolve the gas associated with individual clusters and any impact of stellar feedback. Finally, repeating these calculations using JWST observations sensitive to PAH and small dust grain emission also offers a promising way forward \citep[e.g., expanding on][]{whitmore_empirical_2025}. Such observations are now widely available for this sample \citep[][]{lee_phangs-jwst_2023,chown_polycyclic_2025} and these data have better resolution and sensitivity compared to the current ALMA CO (2-1) maps, though the interpretation of the emission can be more ambiguous \citep[e.g.,][]{leroy_phangs_2023}. 

\begin{acknowledgements}

This work was carried out as part of the PHANGS collaboration.  Based on PHANGS (Physics at High Angular resolution in Nearby Galaxies) observations made with the NASA/ESA Hubble Space Telescope, obtained from MAST (Mikulski Archive for Space Telescopes) at the Space Telescope Science Institute. STScI is operated by the Association of Universities for Research in Astronomy, Inc. under NASA contract NAS 5-26555.  Support for Program number 15654 was provided through a grant from the STScI under NASA contract NAS5-26555.

HH and FB gracefully acknowledge support by Our Dynamic Universe, funded by the Deutsche Forschungsgemeinschaft (DFG, German Research Foundation) under Germany’s Excellence Strategy EXC 3037 – 533607693– Our Dynamic Universe. HH thanks Eric Emsellem for insightful comments and last-minute proof-read. 

AKL gratefully acknowledges support from NSF AST AWD 2205628, JWST-GO-02107.009-A, and JWST-GO-03707.001-A and a Humboldt Research Award. AKL also thanks Liam Dubay for useful conversations.

ER acknowledges the support of the Natural Sciences and Engineering Research Council of Canada (NSERC), funding reference number RGPIN-2022-03499.

AH acknowledges support by the Programme National Cosmology et Galaxies (PNCG) of CNRS/INSU with INP and IN2P3, co-funded by CEA and CNES, and by the Programme National Physique et Chimie du Milieu Interstellaire (PCMI) of CNRS/INSU with INC/INP co-funded by CEA and CNES.


T.G.W gratefully acknowledges support from the UK ALMA Regional Centre (ARC) Node, which is supported by the Science and Technology Facilities Council grant number ST/Y004108/1.

MB acknowledges support by the ANID BASAL project FB210003. This work was supported by the French government through the France 2030 investment plan managed by the National Research Agency (ANR), as part of the Initiative of Excellence of Université Côte d’Azur under reference No. ANR-15-IDEX-01. This research was funded, in whole or in part, by the French National Research Agency (ANR), grant ANR-24-CE92-0044 (project STARCLUSTERS).

LR gratefully acknowledges funding from the DFG through an Emmy Noether Research Group (grant number CH2137/1-1).

ZB, FB and DC gratefully acknowledge the Collaborative Research Center 1601 (SFB 1601 sub-project B3) funded by the Deutsche Forschungsgemeinschaft (DFG, German Research Foundation) – 500700252.

SCOG acknowledges support from the European Research Council via Synergy Grant ``ECOGAL'' (project ID 855130), from the German Excellence Strategy via the Heidelberg Cluster ``STRUCTURES'' (EXC 2181 - 390900948), and from the DFG 
in project GL 668/5-1 (project-ID 585721393).

AU and MQ acknowledge support from the Spanish grants PID2022-138560NB-I00 and PID2025-169300NB-I00, funded by MCIN/AEI/10.13039/501100011033/FEDER, EU.

This work is based on observations made with the NASA/ESA Hubble Space Telescope (program Nos. 15654 and 17126) and NASA/ESA/CSA James Webb Space Telescope (program No. 2107). The data were obtained from the Mikulski Archive for Space Telescopes at the Space Telescope Science Institute, which is operated by the Association of Universities for Research in Astronomy, Inc., under NASA contract 5-26555 for HST and NAS 5-03127 for JWST.

This paper makes use of the following ALMA data, which have been processed as part of the PHANGS--ALMA CO~(2-1) survey: \\
\noindent ADS/JAO.ALMA\#2012.1.00650.S, \linebreak 
ADS/JAO.ALMA\#2013.1.00803.S, \linebreak 
ADS/JAO.ALMA\#2013.1.01161.S, \linebreak 
ADS/JAO.ALMA\#2015.1.00121.S, \linebreak 
ADS/JAO.ALMA\#2015.1.00782.S, \linebreak 
ADS/JAO.ALMA\#2015.1.00925.S, \linebreak 
ADS/JAO.ALMA\#2015.1.00956.S, \linebreak 
ADS/JAO.ALMA\#2016.1.00386.S, \linebreak 
ADS/JAO.ALMA\#2017.1.00392.S, \linebreak 
ADS/JAO.ALMA\#2017.1.00766.S, \linebreak 
ADS/JAO.ALMA\#2017.1.00886.L, \linebreak 
ADS/JAO.ALMA\#2018.1.01321.S, \linebreak 
ADS/JAO.ALMA\#2018.1.01651.S, \linebreak 
ADS/JAO.ALMA\#2018.A.00062.S, \linebreak 
ADS/JAO.ALMA\#2019.1.01235.S, \linebreak 
ADS/JAO.ALMA\#2019.2.00129.S, \linebreak 
ALMA is a partnership of ESO (representing its member states), NSF (USA), and NINS (Japan), together with NRC (Canada), NSC and ASIAA (Taiwan), and KASI (Republic of Korea), in cooperation with the Republic of Chile. The Joint ALMA Observatory is operated by ESO, AUI/NRAO, and NAOJ. The National Radio Astronomy Observatory is a facility of the National Science Foundation operated under cooperative agreement by Associated Universities, Inc.

\end{acknowledgements}

%
%
\bibliographystyle{aa/aa} 
\bibliography{references, ref_2} 

\appendix

\section{Comparison between the ground-based and HST H$\alpha$ observations}
\label{sec:ground_vs_hst}

\begin{figure*}
\centering
\includegraphics[width=0.3\textwidth]{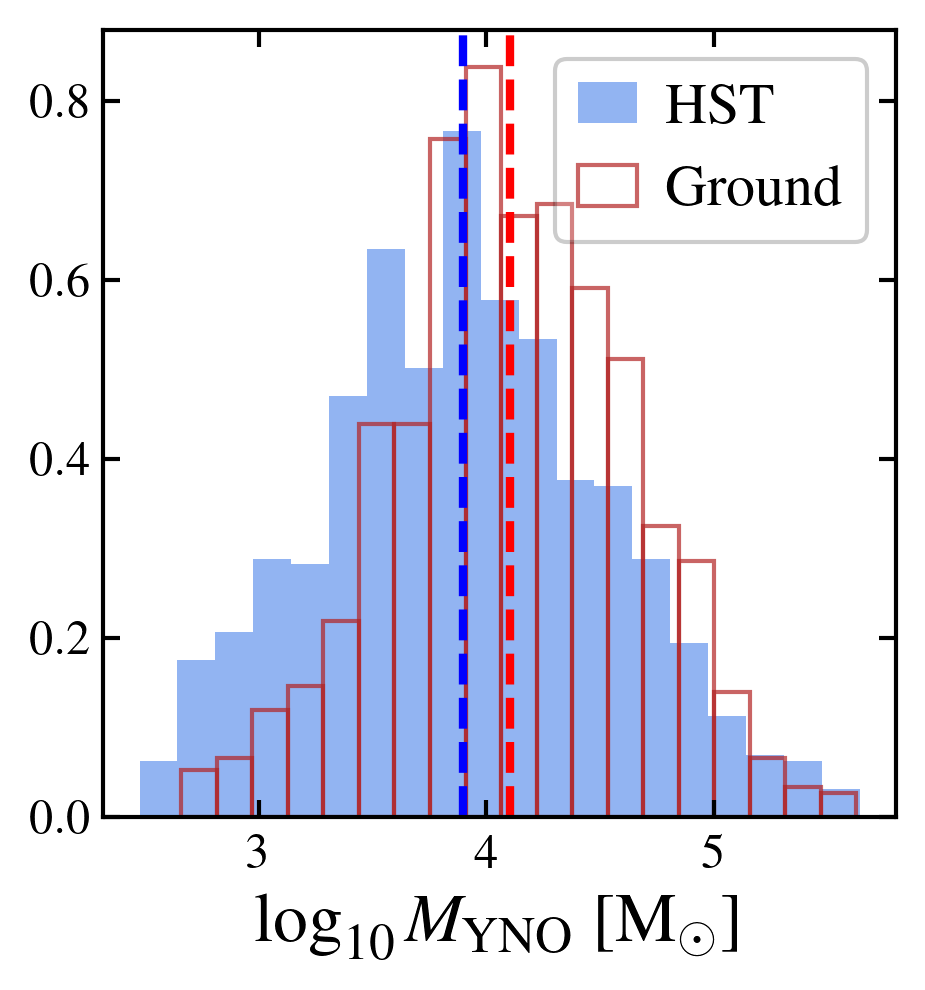}
\includegraphics[width=0.63\textwidth]{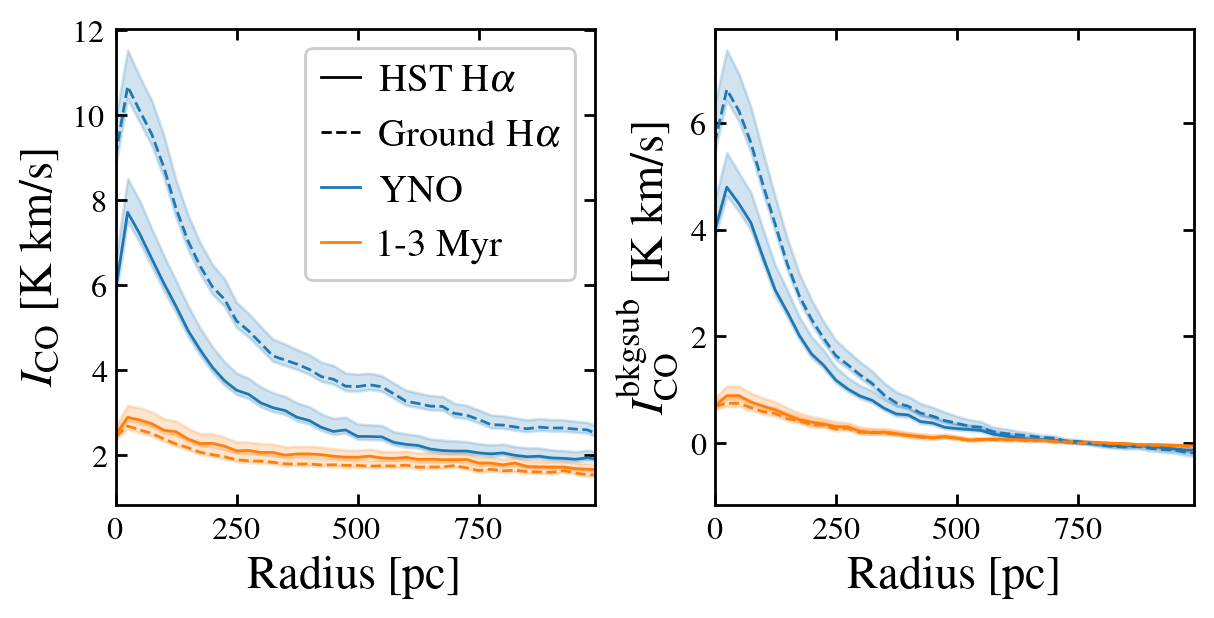}
\caption{(\textit{Left}) The mass distribution for YNO clusters selected from HST (\textit{blue}) and ground-based (\textit{red}) H$\alpha$ data for 16 galaxies with both observations available. The blue and red vertical dashed lines indicate the median cluster mass for HST and ground-based YNO clusters. (\textit{Middle}) The stacked median of $I_{\rm CO}$ for YNO clusters (\textit{blue}) and non-YNO clusters at age of 1--3 Myr (\textit{orange}). The solid and dashed lines indicate HST and ground-based observations. (\textit{Right}) The stacked median of background subtracted intensity, $I_{\rm CO}^{\rm bkgsub}$, for different cluster groups and observations.  }
\label{fig:hst_vs_ground}
\end{figure*}

In our study, we use YNO clusters classified from two narrow-band H$\alpha$ surveys, HST-H$\alpha$ \citep{chandar_2025} and ground-based H$\alpha$ \citep{razza_2026}. These two observations have different spatial resolution and flux sensitivity, and hence can introduce systematic uncertainty in measuring statistical properties of YNO clusters. In Fig. \ref{fig:hst_vs_ground}, we compare the cluster mass distribution and the stacked median $I_{\rm CO}$ and $I_{\rm CO}^{\rm bkgsub}$ for clusters in 16 galaxies that are covered by both observations. We see that ground-based YNO clusters have a median mass $\sim$0.2 dex higher than that of HST YNO clusters (left panel). This likely reflects the different flux thresholds applied when using the two data sets during YNO cluster identification. YNO candidates are first identified based on spatial association with segmented H$\alpha$ clumps in H$\alpha$ maps with diffuse emission removed \citep{thilker_phangs-hst_2025}. Due to the coarser resolution of the ground-based observations ($\sim$1\arcsec), a large fraction of ground-based YNO candidates are falsified by HST morphological criterion. To reduce the false-positive fraction from ground-based only identification, the ground-based flux threshold is more stringent ($\sim$25\% percentile of H$\alpha$ flux for candidates that satisfy criterion from both observations) than the HST flux threshold \citep[$\sim$5\% percentile,][]{thilker_phangs-hst_2025}. Therefore, the YNO catalogue from HST observations include a larger fraction of YNO clusters with faint but significant H$\alpha$ emission, which brings down the median mass of the overall YNO population. 

This different mass sensitivity leads to a difference in the measured distribution of $I_{\rm CO}$ and $I_{\rm CO}^{\rm bkgsub}$ (right panel), since both $I_{\rm CO}^{\rm bkgsub}$ and $I_{\rm CO}^{\rm bkg}$ correlate with the mass of the YNO cluster (Fig. \ref{fig:Idiff_mass_young} and \ref{fig:bkg_vs_mass}). The difference reaches $\sim$2~K~km~s$^{-1}$ in $I_{\rm CO}^{\rm bkgsub}$, which accounts for $\sim$50\% of peak $I_{\rm CO}^{\rm bkgsub}$ for HST YNO clusters. 

The HST-H$\alpha$ based catalogue includes lower mass clusters and $I_{\rm CO}^{\rm bkgsub}$ correlates with $M_{\rm SC}$. We often treat $M_{\rm mol}/M_{\rm SC}$ as a key measurement, and once we normalized by the cluster mass in this way, the bias between classification schemes appears much reduced. The HST YNO clusters have higher $M_{\rm mol}/M_{\rm SC} \approx 104$ than ground-based YNO clusters ($M_{\rm mol}/M_{\rm SC} \approx 87$). Therefore we mostly expect that the galaxies with HST H$\alpha$ just extend the trends seen in the ground based H$\alpha$ to better sensitivity.

In principle, the inclusion of clusters not identified as YNOs could also affect the 1--3 Myr bin of the YNO-clusters, but their CO intensities for non-YNO clusters appear less affected (orange in middle and right panel of Fig. \ref{fig:hst_vs_ground}). 

\section{Separating YNO clusters by age}
\label{sec:yno_age_test}

\begin{figure*}
\centering
\includegraphics[width=0.9\textwidth]{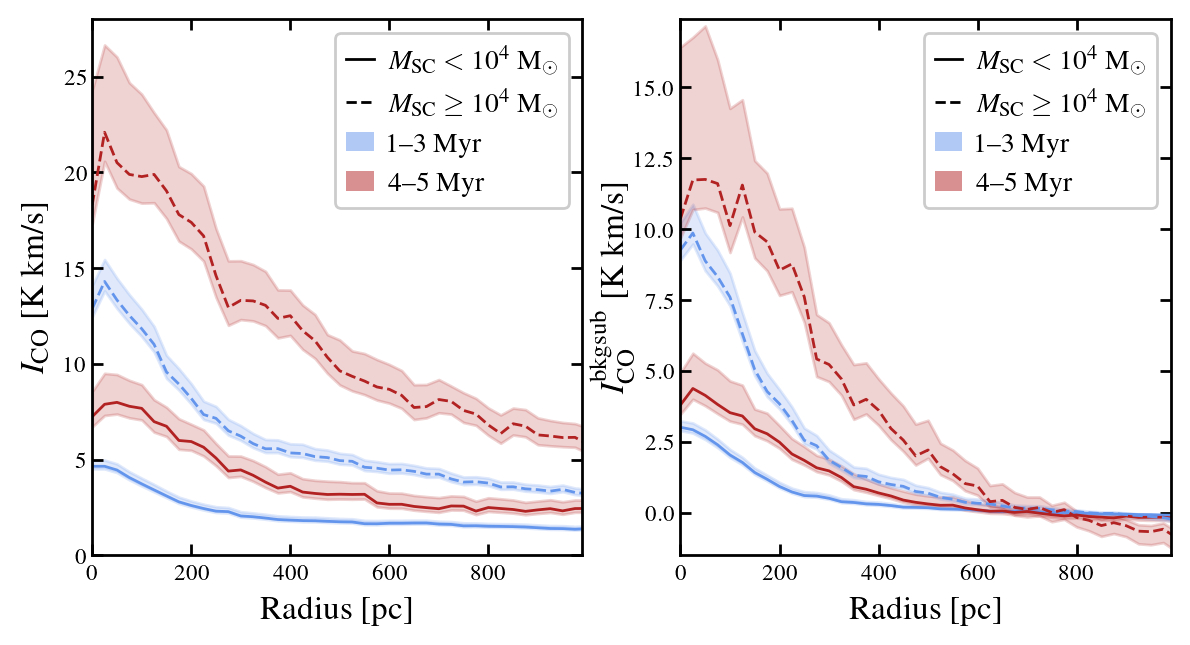}
\caption{The stacked median of the radial profile of (\textit{left}) $I_{\rm CO}$ and (\textit{right}) $I_{\rm CO}^{\rm bkgsub}$ for YNO clusters. The solid and dashed lines indicate clusters with mass lower and higher than 10$^4$ \solarmass. Blue and red colours indicate cluster at 1--3 Myr and 4--5 Myr, respectively. We see that YNO clusters of 4--5 Myr generally shows a even stronger association with CO(2-1) intensity peaks than YNO clusters of 1--3 Myr.   }
\label{fig:yno_intensity_profl}
\end{figure*}

The physical properties of YNO clusters are determined using SED fitting with a specific ``YNO'' template grid \citep{thilker_phangs-hst_2025}. Compared to general grids, YNO grids limit cluster age within  1--5 Myr to reduce the age-reddening degeneracy in SED fitting, as many clusters appear as globular clusters without reddening correction. On  the stellar track in \textit{UVBI} colour-colour diagram, clusters at age of 1--3 Myr and 4--5 Myr occupy different positions and can be separated. However, it is not clear if YNO clusters of these two age ranges can be well separated with significant reddening correction. Here we test if YNO clusters of different age groups also show the same co-evolutionary trend with CO that we see for non-YNO clusters (Fig. \ref{fig:cluster_co_intensity}). 

In Fig. \ref{fig:yno_intensity_profl}, we show the CO intensity profile comparison between YNO clusters of 1--3 Myr and 4--5 Myr. For both low- ($M<10^4$ \solarmass) and high-mass ($M\geq10^4$ \solarmass) clusters, we see that YNO clusters of 4--5 Myr have higher $I_{\rm CO}$ and $I_{\rm CO}^{\rm bkgsub}$ than the supposedly younger population of age 1--3 Myr. This inverse trend could potentially be related to the cluster mass difference between the two age groups, as YNO clusters of 4--5 Myr are more massive (Fig. \ref{fig:mass_vs_age}). However, after normalized by cluster mass, we still find a higher $M_{\rm mol}/M_{\rm SC}$ ratio  for clusters of 4--5 Myr ($\sim$101) than the ratio of 1--3 Myr ($\sim$78). Overall, this suggests that the SED fitting is unable to separate YNO clusters of these two age groups. We proceed in the main text grouping all YNOs into a single category. 

\section{Impact of resolution on the stacked CO profiles}
\label{sec:co_res}

\begin{figure}
    \centering
    \includegraphics[width=0.45\textwidth]{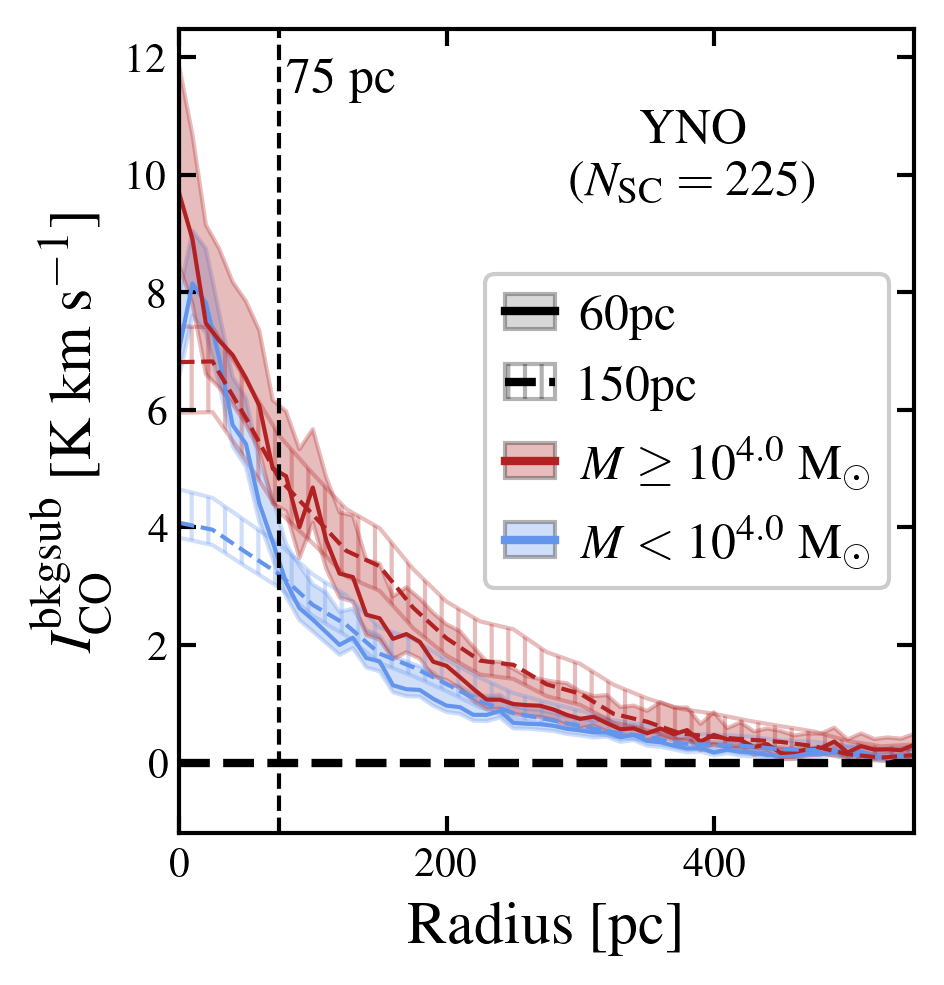}
    \vspace{-1\baselineskip}
    \caption{The stacked median of radial profiles of $I_{\rm CO}^{\rm bkgsub}$ for YNO clusters covered by both 60 pc and 150 pc resolution CO observations. Lines indicate stacked median at 60 pc (solid) and 150 pc (dashed) resolution. The different colours indicate clusters lower (\textit{blue}) and higher (\textit{red}) than 10$^4$ \solarmass. The vertical dashed line indicates the half width of the 150 pc beam, which is roughly the scale beyond which the profiles of two resolutions start to converge.  }
    \label{fig:yno_profl_res}
\end{figure}

Our adopted resolution of 150~pc is larger than the typical GMC size found in Milky Way \citep[often 10--40 pc;][]{heyer_molecular_2015}, and hence we might expect beam dilution to affect our measured CO intensity. To quantify the impact of resolution, we construct radial profiles of $I^{\rm bkgsub}_{\rm CO}$ at two spatial resolutions, 60~pc and 150~pc, for clusters in a sub-sample of seven galaxies with both CO maps available, as shown in Fig. \ref{fig:yno_profl_res}. For both high and low YNO clusters we find that the higher resolution data show larger peak $I_{\rm CO}^{\rm bkgsub}$. In both cases the difference is $\sim$ 3 K~km~s$^{-1}$ in the peak intensity. 
This further supports that YNO clusters are spatially associated with high-density regions of molecular clouds where massive cluster formation is ongoing.
The radial profiles of the 60~pc and 150~pc resolution data begin to converge beyond the half width at half maximum of the lower-resolution beam (HWFM=75~pc). 
This convergence suggests that our background subtraction method works consistently at scales above resolution limit for different CO-resolution maps. 

We also attempted a similar comparison for the non-YNO clusters. However, due to the limited sample size of 60~pc resolution CO maps ($\sim$1/5 of the number of our main sample, Table \ref{tab:Idiff_results}), we generally see flat radial profiles at both resolutions even for the youngest non-YNO clusters (1--3 Myr old). A larger sample of high-resolution CO maps will be needed to probe resolved clouds and later cluster evolutionary stages.  

\section{Impact from age uncertainty}
\label{sec:age_uncertainty}

\begin{figure}
    \centering
    \includegraphics[width=0.45\textwidth]{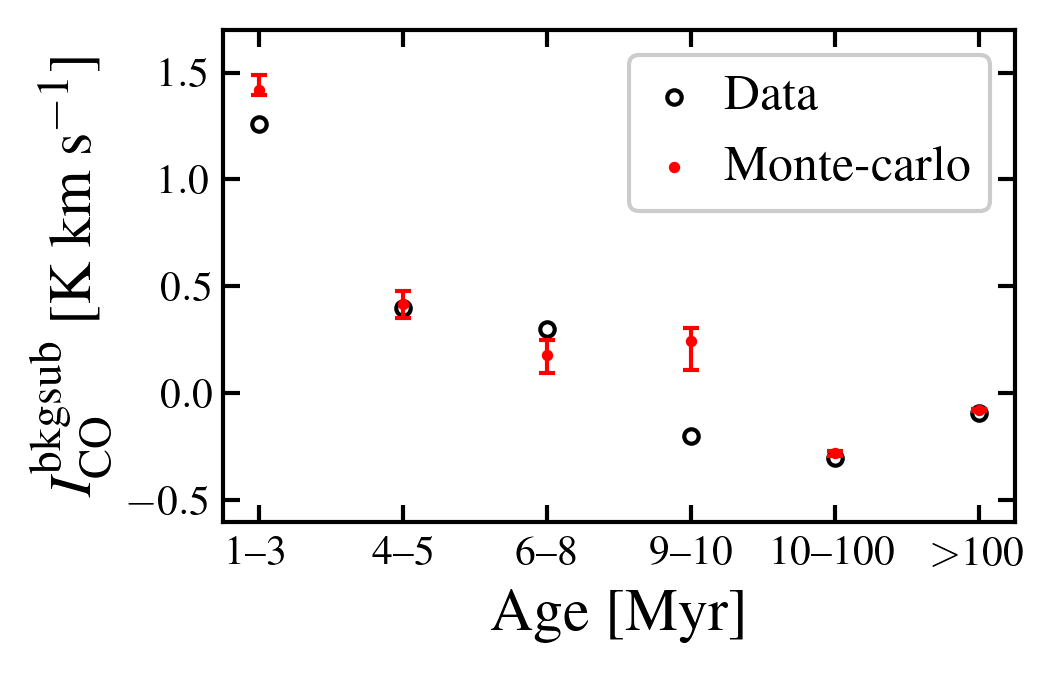}
    \caption{Monte Carlo estimate of the age uncertainty impact on the measured $I_{\rm CO}^{\rm bkgsub}$ of different age groups. The empty black circles indicate measurements from real data while the red error bar indicates the median and 16th-84th range from the Monte-Carlo simulation.  }
    \label{fig:age_monte_carlo}
\end{figure}

Age uncertainty for individual clusters can impact our results. Very young clusters (age $\leq$ 3 Myr) have a typical age uncertainty of 0.2 -- 0.4 dex, which includes photometric uncertainty and model degeneracy. Older clusters (several to several hundred Myr old) can have a maximal age uncertainty of 0.15 dex \citep{thilker_phangs-hst_2025}. 

To quantify the impact of these age uncertainties we perform a Monte Carlo simulation. For each cluster, we draw lower and upper limits of the age from the PHANGS-HST catalogue \citep[derived based on the posterior distribution from SED fitting,][]{thilker_phangs-hst_2025}. Then we draw a new age sampled across the range between the given lower and upper limit following a uniform distribution. We then group the clusters based on the Monte Carlo-generated new ages into our previously used age bins and make the same measurement of $I_{\rm CO}^{\rm bkgsub}$ for each age group. We repeat this process 100 times, and calculate the median and 16th-84th scatter of the measured quantities. 

In Fig.~\ref{fig:age_monte_carlo}, we show the simulated $I_{\rm CO}^{\rm bkgsub}$ for different age groups and corresponding measured values. We see that Monte Carlo-simulated quantities follow a similar age trend as observed values. This suggests that our observed cluster-gas co-evolution trend is robust against age uncertainties. 

We note that the expected quantity from Monte Carlo simulation can deviate from real value by more than 16th-84th scatter range. This is generally due to the fact that the age uncertainty range is asymmetric. With our simplified uniform sampling method within the age uncertainty range, our expected age median can significantly deviate from the catalogue-given age of the maximal likelihood for individual clusters. This conservative approach tends to overestimate the age uncertainty impact, and hence further demonstrates that our results are robust against this uncertainty. 

The deviation between the observed quantity and the Monte-Carlo modelled results suggest possible contamination of certain age groups. Specifically, clusters at 1--3 Myr show lower observed $I_{\rm CO}^{\rm bkgsub}$ than the modelled results, possibly due to older clusters with less associated CO/molecular gas coming into this age group. From modelling, the median number of clusters of this age group is 2126 (compared to 2359 from real observations), which suggests a contamination fraction of $\sim$10\%. Similarly, clusters of 9--10 Myr from observation might also be contaminated by older clusters of 10--100 Myr, and hence show lower observed values of $I_{\rm CO}^{\rm bkgsub}$. 

\section{Summary of individual galaxies}
\label{sec:gal_indvd}

\begin{figure*}
\centering
\includegraphics[width=0.9\textwidth]{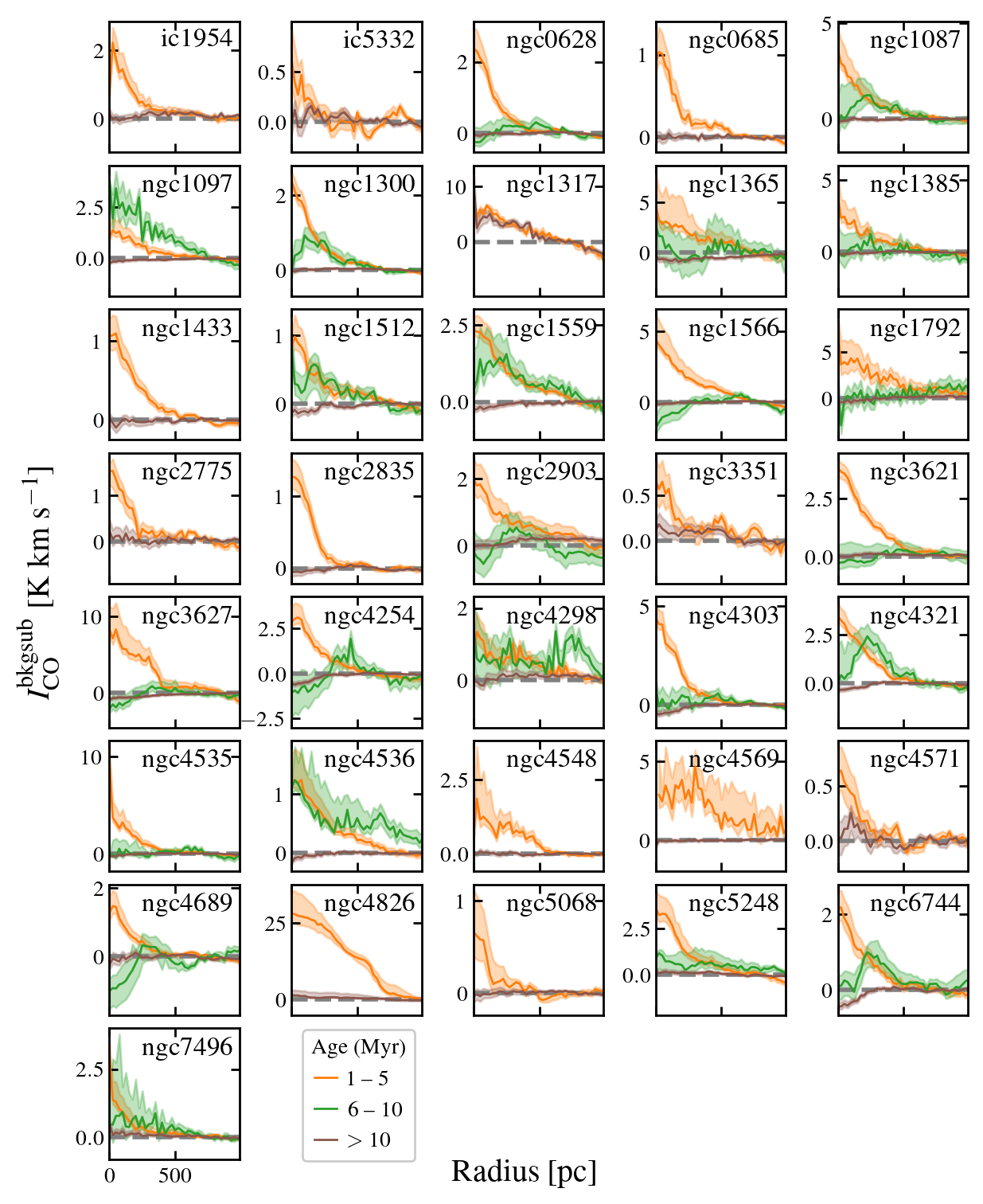}
\caption{For each individual galaxy in our sample, the stacked median $I_{\rm CO}^{\rm bkgsub}$ radial profile for clusters with mass greater than 10$^{3.5}$ \solarmass, and at age of 1--5 Myr (\textit{orange}), 6--10 Myr (\textit{green}) and $>$ 10 Myr (\textit{brown}). The shaded regions indicate the uncertainty in the median derived from error propagation. For the 1--5 Myr group, we combine YNO and non-YNO clusters. We exclude age groups with less than $10$ clusters.  }
\label{fig:stacking_bkgsub_indvd}
\end{figure*}

\begin{table*}
\caption{\label{tab:gal_summary} Summary of star cluster and molecular gas properties of individual galaxies}
\centering
\renewcommand{\arraystretch}{1.4}
\setlength{\tabcolsep}{3pt}
\begin{tabularx}{\textwidth}{ccccccccccc}
\hline\hline 
Galaxy & Morphology & $\log_{10}\Sigma_{\rm SFR}$ & $\log_{10}\Sigma_{\rm mol}$ & \multicolumn{3}{c}{Young clusters (age$\leq$10 Myr)} &  \multicolumn{3}{c}{YNO clusters} \\
 & &  &  & $N_{\rm SC}$ & $\log_{10} M_{\rm SC}$ & $I^{\rm bkg}_{\rm CO}$  & $N_{\rm YNO}$ & $\log_{10} M_{\rm YNO}$ & $I_{\rm CO}^{\rm bkgsub}$ & $M_{\rm mol}/M_{\rm YNO}$ \\
  & & [M$_{\odot}$~yr$^{-1}$~kpc$^{-2}$] & [M$_{\odot}$~kpc$^{-2}$] & & [\solarmass] & [K~km~s$^{-1}$] & & [\solarmass] & [K~km~s$^{-1}$] &  \\
(1) & (2) & (3) & (4) & (5) & (6) & (7) & (8) & (9) & (10) & (11) \\
\hline
ic1954 & Sb & -2.0 & 7.2 & 54 & 3.8 & $1.6\pm0.5$ & 19 & 3.8 & $5_{-2}^{+0.9}$ & $59_{-10}^{+17}$ \\
ic5332 & SABc & -2.3 & 6.0 & 14 & 3.7 & $0.31\pm0.09$ & 3 & 3.9 & $1.6_{-0.5}^{+0.6}$ & $37_{-10}^{+11}$ \\
ngc0628 & Sc & -1.7 & 7.5 & 144 & 3.8 & $1.78\pm0.05$ & 64 & 3.8 & $4.6_{-0.7}^{+0.7}$ & $100_{-10}^{+15}$ \\
ngc0685 & Sc & -2.6 & 6.6 & 60 & 4.1 & $0.26\pm0.04$ & 9 & 4.3 & $2_{-0.8}^{+1}$ & $24_{-10}^{+11}$ \\
ngc1087 & Sc & -1.7 & 7.4 & 140 & 4.1 & $2\pm0.3$ & 39 & 4.3 & $10_{-3}^{+5}$ & $68_{-10}^{+15}$ \\
ngc1097 & SBb & -0.9 & 8.1 & 150 & 3.9 & $1.4\pm0.1$ & 0 &  &  &  \\
ngc1300 & Sbc & -2.4 & 7.0 & 237 & 3.8 & $0.79\pm0.05$ & 86 & 3.9 & $3.8_{-0.4}^{+0.5}$ & $79_{-10}^{+11}$ \\
ngc1317 & SABa & -1.6 & 7.6 & 15 & 4.2 & $9\pm3$ & 7 & 4.2 & $8_{-0.9}^{+1}$ & $68_{-20}^{+57}$ \\
ngc1365 & Sb & -0.5 & 8.6 & 77 & 4.3 & $3.4\pm0.4$ & 23 & 4.2 & $18_{-7}^{+8}$ & $112_{-50}^{+82}$ \\
ngc1385 & Sc & -1.5 & 7.4 & 104 & 4.2 & $2.4\pm0.4$ & 40 & 4.3 & $7_{-1}^{+1}$ & $52_{-20}^{+19}$ \\
ngc1433 & SBa & -2.0 & 7.2 & 85 & 3.8 & $0.49\pm0.06$ & 16 & 3.8 & $3.1_{-0.6}^{+0.6}$ & $44_{-10}^{+19}$ \\
ngc1512 & Sa & -2.0 & 7.0 & 53 & 3.7 & $0.5\pm0.06$ & 6 & 3.7 & $1.2_{-0.8}^{+2}$ & $44_{-10}^{+21}$ \\
ngc1559 & SBc & -1.4 & 7.6 & 376 & 4.2 & $3\pm0.3$ & 84 & 4.5 & $8_{-1}^{+1}$ & $28_{-5}^{+7}$ \\
ngc1566 & SABb & -1.1 & 7.9 & 231 & 4.1 & $3.2\pm0.3$ & 63 & 4.2 & $20_{-5}^{+5}$ & $138_{-30}^{+34}$ \\
ngc1792 & Sbc & -1.5 & 7.8 & 148 & 4.5 & $9.4\pm0.9$ & 41 & 4.6 & $24_{-4}^{+3}$ & $69_{-10}^{+21}$ \\
ngc2775 & Sab & -2.2 & 7.1 & 59 & 4.0 & $2.6\pm0.1$ & 0 &  &  &  \\
ngc2835 & Sc & -1.7 & 6.9 & 83 & 3.8 & $0.45\pm0.08$ & 35 & 3.7 & $2.6_{-0.5}^{+0.4}$ & $48_{-10}^{+11}$ \\
ngc2903 & Sbc & -1.5 & 7.6 & 302 & 4.0 & $5.6\pm0.3$ & 8 & 4.5 & $12_{-2}^{+3}$ & $64_{-10}^{+30}$ \\
ngc3351 & Sb & -1.6 & 7.3 & 41 & 3.8 & $1.19\pm0.07$ & 6 & 3.6 & $1.5_{-0.8}^{+1}$ & $52_{-30}^{+50}$ \\
ngc3621 & SBcd & -1.7 & 7.4 & 211 & 3.7 & $4.6\pm0.5$ & 96 & 3.8 & $6_{-0.7}^{+0.8}$ & $126_{-20}^{+20}$ \\
ngc3627 & Sb & -1.3 & 7.9 & 165 & 4.1 & $7.1\pm0.4$ & 51 & 4.3 & $22_{-5}^{+12}$ & $112_{-30}^{+44}$ \\
ngc4254 & Sc & -1.1 & 8.3 & 414 & 4.0 & $7\pm1$ & 118 & 4.1 & $9_{-1}^{+1}$ & $74_{-8}^{+11}$ \\
ngc4298 & Sc & -2.1 & 7.4 & 115 & 4.0 & $3.6\pm0.4$ & 8 & 4.1 & $6_{-2}^{+0.7}$ & $53_{-20}^{+19}$ \\
ngc4303 & Sbc & -1.1 & 8.0 & 507 & 4.2 & $4.1\pm0.2$ & 110 & 4.3 & $12_{-2}^{+3}$ & $83_{-10}^{+16}$ \\
ngc4321 & SABb & -1.7 & 7.6 & 245 & 4.0 & $3.6\pm0.1$ & 52 & 4.0 & $8_{-1}^{+0.9}$ & $89_{-20}^{+22}$ \\
ngc4535 & Sc & -2.1 & 7.2 & 80 & 4.2 & $1.8\pm0.1$ & 22 & 4.5 & $12_{-5}^{+4}$ & $83_{-20}^{+14}$ \\
ngc4536 & SABb & -1.5 & 7.3 & 126 & 4.0 & $0.67\pm0.08$ & 4 & 4.6 & $5_{-3}^{+38}$ & $27_{-10}^{+77}$ \\
ngc4548 & Sb & -2.6 & 7.0 & 36 & 3.9 & $1\pm0.1$ & 5 & 4.5 & $12_{-1}^{+3}$ & $74_{-30}^{+75}$ \\
ngc4569 & Sab & -2.2 & 7.3 & 29 & 4.5 & $5.5\pm0.9$ & 2 & 4.6 & $12_{-1}^{+1}$ & $54_{-10}^{+16}$ \\
ngc4571 & Sc & -2.5 & 6.9 & 41 & 3.8 & $1.2\pm0.1$ & 1 & 4.0 & $3.7_{-0.9}^{+0.8}$ & $65_{-20}^{+19}$ \\
ngc4689 & Sc & -2.5 & 6.9 & 100 & 4.0 & $1.7\pm0.2$ & 14 & 4.3 & $3.9_{-0.8}^{+0.9}$ & $67_{-20}^{+16}$ \\
ngc4826 & SABa & -1.9 & 7.4 & 12 & 4.1 & $2.5\pm0.7$ & 4 & 4.0 & $30_{-8}^{+23}$ & $582_{-300}^{+311}$ \\
ngc5068 & Sc & -1.9 & 7.0 & 33 & 3.8 & $0.41\pm0.09$ & 11 & 3.6 & $1.5_{-0.6}^{+1}$ & $52_{-20}^{+19}$ \\
ngc5248 & SABb & -1.4 & 7.9 & 137 & 4.1 & $3.4\pm0.3$ & 33 & 4.3 & $11_{-2}^{+1}$ & $86_{-20}^{+19}$ \\
ngc6744 & Sbc & -2.1 & 7.0 & 63 & 3.8 & $1.5\pm0.1$ & 12 & 3.7 & $5_{-2}^{+2}$ & $120_{-30}^{+42}$ \\
ngc7496 & Sb & -1.6 & 7.3 & 81 & 4.0 & $1.1\pm0.1$ & 21 & 4.1 & $5.2_{-0.9}^{+2}$ & $80_{-30}^{+31}$ \\
\hline
\end{tabularx}
\tablefoot{Columns: (1) Galaxy name; (2) morphological classification; (3) star formation rate surface density within the stellar effective radius; (4) global molecular gas surface density within the stellar effective radius;
(5) number of clusters younger than 10 Myr; (6) median cluster stellar mass, $M_{\rm SC}$, for clusters younger than 10 Myr (7) median and uncertainty in the median for the background subtracted CO intensity associated with clusters younger than 10 Myr (8) The number of clusters classified as young nebular objects (YNO) (9) The median mass of YNO clusters (10) The median and the median uncertainty of the background subtracted CO intensity (11) The median and the uncertainty of molecular gas mass to cluster mass ratio. 
}
\end{table*}

We summarize the cluster properties and the associated CO intensity measurements for individual galaxies in Table \ref{tab:gal_summary}. For each galaxy, we exclude clusters with mass below 10$^{3.5}$ \solarmass to ensure a common completeness level. We also include the galaxy morphology and molecular gas and SFR surface density within the stellar effective radius of the galaxy  \citep{leroy_phangs-alma_2021}.

We plot the stacked median $I_{\rm CO}^{\rm bkgsub}$ profiles of each galaxy in Fig. \ref{fig:stacking_bkgsub_indvd}. Due to the limited sample size of clusters in individual galaxies, we define coarser age bins compared to what we use for the overall sample. Here the clusters are stratified into three age groups: 1-5~Myr, 5-10~Myr and $>$10 Myr. The first age group includes both YNO and non-YNO clusters. We do not show the intensity profiles if the given age group has less than ten clusters. 

In general, we find that most galaxies follow a similar evolutionary trend as the entire sample. Young clusters of age 1--5 Myr show strong central CO concentrations while clusters older than 10 Myr generally have flat profiles (except for NGC 1317). Clusters with ages of 6--10 Myr show more complex behaviour. This might be due to the relatively small sample size for clusters within this range; there are $\sim$1/5 as many 6--10 Myr clusters compared to 1--5 Myr old ones and $\sim$1/20 as many 6--10 Myr clusters compared to clusters older than 10 Myr (see Table \ref{tab:Idiff_results}). 

Many galaxies (e.g. NGC 1087, NGC 1300, NGC 1512, NGC 1559, NGC 4321, NGC 4689, and NGC 6744) show a bump at a cluster-centric radius of 200--300~pc within cluster age range of 6-10 Myr. This feature is analogous to what we see in the stacked $I_{\rm CO}^{\rm bkgsub}$ profile for all massive clusters with age at 6--8 Myr (Fig. \ref{fig:stack_Idiff_mass}), and might reflect the signature of stellar feedback in re-shaping the surrounding molecular gas structure. 

\section{Star cluster distribution in NGC 628}

\begin{figure*}
    \centering
    \gridline{
    \fig{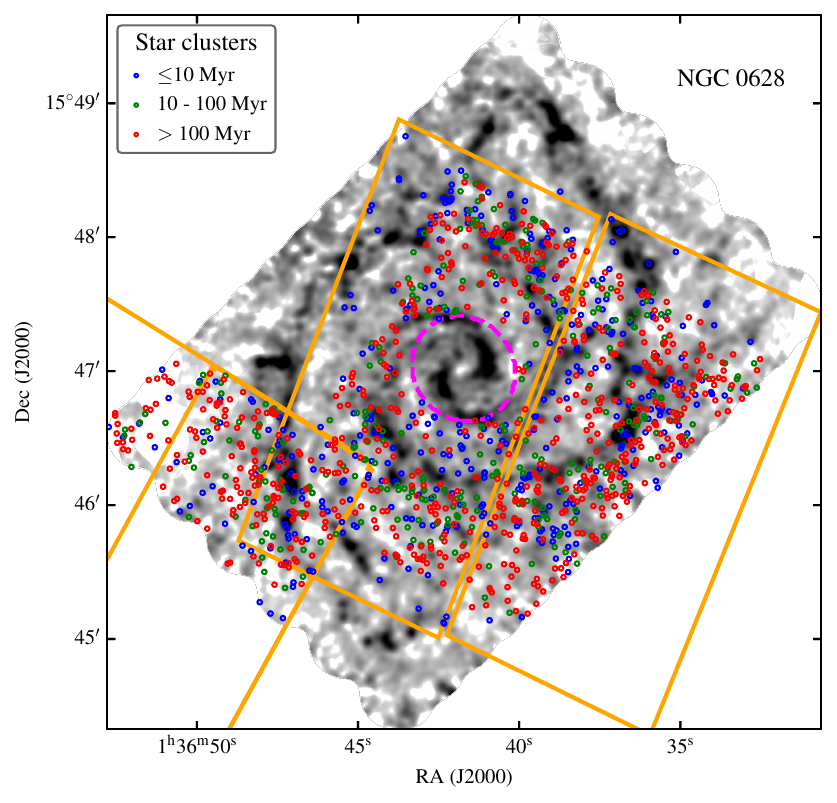}{0.9\textwidth}{}
    }
    \vspace{-2\baselineskip}
    \caption{The HST star clusters \citep{maschmann_phangs-hst_2024, thilker_phangs-hst_2025} overlaid on the PHANGS-ALMA CO~(2-1) maps for NGC 628. The blue, green and red circles indicate clusters of 1--10 Myr, 10--100 Myr and $>$100 Myr old. The orange rectangles indicate the HST field of view (FOV). The magenta dashed circle indicates the central region where we mask out star clusters from our analyses. A large fraction of clusters younger than 10 Myr are in or near the spiral arms seen in CO, while old clusters (age $>$ 100 Myr) have a more smooth spatial distribution.}
    \label{fig:ngc628_maps}
\end{figure*}

In Fig. \ref{fig:ngc628_maps}, we show an example of the cluster distribution overlaid on the CO~(2-1) maps at 150 pc resolution in NGC 628. We stratify clusters in into three age groups: 1-10 Myr, 10-100 Myr and $>$100 Myr. Clusters younger than 10 Myr have a spatial distribution that traces the CO spiral arms. 
Clusters with ages 10--100 Myr generally reside near but not in spiral arms, and clusters older than 100 Myr have a more random distribution. In both cases the majority of these older clusters are found in the interarm regions. 
%
These different spatial distributions manifest as the correlation between cluster age and $I_{\rm CO}^{\rm bkg}$ in Fig. \ref{fig:bkg_vs_mass} and \S \ref{subsec:profile_outer}. The stronger central dip in the $I_{\rm CO}^{\rm bkgsub}$ profile for intermediate age (10-100 Myr) clusters compared older clusters ($>$100 Myr, Fig. \ref{fig:profl_dip}) might also be explained by their different proximity to the large-scale CO structures (i.e. the arms).

\end{document}